\documentclass[acmsmall]{acmart}

\AtBeginDocument{%
  \providecommand\BibTeX{{%
\normalfont B\kern-0.5em{\scshape i\kern-0.25em b}\kern-0.8em\TeX}}}
\usepackage[htt]{hyphenat}
\usepackage{pifont}% http://ctan.org/pkg/pifont
\usepackage[frozencache,cachedir=.]{minted}% for arXiv
\usepackage{listings} % Minted doesn't work on tabularx
\usepackage[scaled=0.85]{DejaVuSansMono}
\usepackage{multirow}
\usepackage{graphicx}
\usepackage{threeparttable}
\usepackage{booktabs}
\usepackage{tabularx}
\usepackage{subcaption}
\usepackage{float}
\usepackage{xcolor}
\usepackage{soul}   % For \hl (highlighting)
\usepackage{enumitem}
\usepackage{balance}
\usepackage{acronym}
\usepackage{glossaries}
\newacronym{mps}{MPS}{Matrix Product States}
\newacronym{msr}{MSR}{mining software repositories}
\newacronym{nisq}{NISQ}{Noisy Intermediate-Scale Quantum}
\newacronym{qc}{QC}{Quantum Computing}
\newacronym{qosf}{QOSF}{Quantum Open Source Foundation}
\newacronym{qpu}{QPU}{Quantum Processing Unit}
\newacronym{qse}{QSE}{Quantum Software Engineering}
\newacronym{pr}{PR}{pull request} %% Acronyms Management

\usepackage{array}

\usepackage{subcaption}
\usepackage{adjustbox}
\usepackage{caption}
\usepackage{url}

\usepackage{algorithmic}
\usepackage{graphicx}
\usepackage{textcomp}

\usepackage{tikz}
\usepackage{tikz-qtree}
\usetikzlibrary{trees}
\usetikzlibrary{positioning}
\usepackage{wrapfig}
\usepackage[framemethod=tikz]{mdframed}

\setminted{
  breaklines=true,
  breakanywhere=false,
  fontsize=\small,
  bgcolor=black!2,
  autogobble=true,
  python3=true,
  mathescape=false,
  texcomments=false,
  numbers=none,
  numbersep=5pt,
  xleftmargin=0pt
}

\setminted[diff]{
  breaklines=true,
  breakanywhere=false,
  fontsize=\small,
  bgcolor=black!2,
  autogobble=true,
  numbers=none,
  numbersep=5pt,
  xleftmargin=0pt
}

\setminted[text]{
  breaklines=true,
  breakanywhere=false,
  breaksymbolleft={}
}

\setmintedinline{
  bgcolor={},
  fontsize=\small
}

\definecolor{cmarkcolor}{RGB}{21, 164, 64}
\definecolor{xmarkcolor}{RGB}{177, 0, 4}

\usepackage[most]{tcolorbox}
\usepackage{csquotes}
\newtcolorbox{myquote}[1][]{%
  colback=black!5,
  colframe=black!5,
  notitle,
  sharp corners,
  borderline west={2pt}{0pt}{black!80!black},
  enhanced,
  breakable,
  top=0.5pt,
  bottom=0.5pt
}

\newenvironment{example}[2][]{%
  \begin{mdframed}[
      linecolor=blue!60,
      linewidth=3pt,
      topline=false,
      rightline=false,
      bottomline=false,
      backgroundcolor=gray!5,
      skipabove=0pt,
      skipbelow=0pt,
      innertopmargin=5pt,
      innerbottommargin=5pt,
      innerleftmargin=5pt,
      innerrightmargin=5pt,
      frametitle={#2},
      frametitlefont=\bfseries\color{blue!60!black},
      frametitlebackgroundcolor=gray!5,
      frametitleaboveskip=0pt,
      frametitlebelowskip=0pt,
      #1
    ]%
    \footnotesize
    \lstset{basicstyle=\footnotesize\ttfamily}
  }{%
  \end{mdframed}%
}

\usepackage{listings}
\definecolor{todocolor}{RGB}{255, 180, 180}

\newcommand{\code}[1]{\lstinline`#1`}

\copyrightyear{2026}
\acmYear{2026}
\setcopyright{cc}
\setcctype{by}

\acmJournal{PACMSE}

\author{Krishna Upadhyay}
\orcid{0009-0002-6854-3186}
\affiliation{%
  \institution{Louisiana State University}
  \city{Baton Rouge}
  \state{LA}
  \country{USA}}
\email{kupadh4@lsu.edu}

\author{Moshood A. Fakorede}
\orcid{0009-0003-2057-9865}
\affiliation{%
  \institution{Louisiana State University}
  \city{Baton Rouge}
  \state{LA}
  \country{USA}}
\email{mfakor1@lsu.edu}

\author{Umar Farooq}
\orcid{0000-0001-7229-9847}
\affiliation{%
  \institution{Louisiana State University}
  \city{Baton Rouge}
  \state{LA}
  \country{USA}}
\email{ufarooq@lsu.edu}

\begin{document}

\title{Understanding Bugs in Quantum Simulators: An Empirical Study}

\newcommand{\stitle}[1]
{\noindent\textup{\textbf{#1}}}
\newcommand\myNum[1]{\emph{{(#1)}}}
\newcommand\todou[1]{\textcolor{red}{\textit{Umar: #1}}}
\newcommand\todok[1]{\textcolor{blue}{\textit{Krishna: #1}}}
\newcommand\todov[1]{\textcolor{blue}{\textit{Vinaik: #1}}}

\begin{abstract}
  Quantum simulators are a foundational component of the quantum software ecosystem. They are widely used to develop and debug quantum programs, validate compiler transformations, and support empirical claims about correctness and performance.
  In the absence of large-scale quantum hardware, simulator outputs are often treated as ground truth for algorithm development and system evaluation.
  However, quantum simulators also introduce unique implementation challenges. They must faithfully emulate quantum behavior while executing on classical hardware, requiring complex representations of quantum state evolution, operator composition, and noise modeling.
  Yet, we still lack a large-scale and in-depth study of failures in quantum~simulators.

  To bridge this gap, this work presents a comprehensive empirical study of bugs in widely used open-source quantum simulators. We analyze 394 confirmed bugs from 12 simulators and manually categorize them based on root causes, failure manifestations, affected components, and discovery mechanisms.
  Our study reveals several key findings. First, bug discovery is largely user-driven, with most crashes, exceptions, and resource-related failures not detected by automated testing and identified after deployment. Second, logical correctness failures are widespread and often silent, producing plausible but incorrect outputs without triggering crashes or explicit error signals. Third, many critical failures originate in classical simulator infrastructure, such as memory management, indexing, configuration, and dependency compatibility, rather than in core quantum execution logic. These findings provide new insights into the reliability challenges of quantum simulators and highlight opportunities to improve testing and validation practices in the quantum software ecosystem.

\end{abstract}

\begin{CCSXML}
  <ccs2012>
  <concept>
  <concept_id>10011007.10011006.10011072</concept_id>
  <concept_desc>Software and its engineering~Software libraries and repositories</concept_desc>
  <concept_significance>500</concept_significance>
  </concept>
  <concept>
  <concept_id>10002944.10011123.10010912</concept_id>
  <concept_desc>General and reference~Empirical studies</concept_desc>
  <concept_significance>300</concept_significance>
  </concept>
  </ccs2012>
\end{CCSXML}

\ccsdesc[500]{Software and its engineering~Software libraries and repositories}
\ccsdesc[300]{General and reference~Empirical studies}

\keywords{quantum computing simulators, software bugs, empirical study}

\maketitle
\section{Introduction}
\label{sec:intro}

Quantum computing introduces a new computational paradigm that offers asymptotic advantages for problems such as molecular simulation, combinatorial optimization, and cryptographic analysis~\cite{Bravyi22}. However, current devices remain limited by qubit count, noise, and error rates, which is defined as the \gls*{nisq} era~\cite{Li19}.
As a result, quantum simulators have become indispensable infrastructure.
They execute quantum programs on classical hardware and serve as the primary environment for developing algorithms, validating compilers, testing optimizations, and benchmarking emerging quantum devices.

Quantum software stacks rely heavily on simulators. Nearly all major quantum frameworks provide one or more simulators as part of their core distribution, such as Qiskit's Aer~\cite{qiskit-aer}, Cirq's QSim~\cite{qsim}, PennyLane's Lightning~\cite{pennylane-lightning}, and standalone tools such as Qulacs~\cite{qulacs} and Qrack~\cite{qrack}, and many users interact exclusively with simulators during development.
In the absence of large-scale fault-tolerant hardware, simulator outputs serve as the primary reference for algorithm validation,  guiding algorithm design decisions, validating transformations, and shaping empirical claims about correctness and performance~\cite{ibm_quantum_simulators, aws_braket_testing, paltenghi2024surveytestinganalysisquantum}. Consequently, failures in simulators can have far-reaching effects across the quantum software ecosystem~\cite{Paltenghi22}.

Despite their importance, the reliability of quantum simulators remains poorly understood.
The implementation of quantum simulators entails a complex mapping of quantum mechanics onto classical architectures.
This introduces unique challenges that differ from those of conventional software systems.
Unlike many infrastructure failures that manifest as crashes or explicit errors, \emph{simulator bugs can produce plausible but incorrect results}, silently violating fundamental correctness properties. Such failures are particularly difficult to detect and can mislead users without warning.

\begin{table}[]
  \centering
  \footnotesize
  % This forces \lstinline (and thus \code command) to use \footnotesize
  % and the typewriter (\ttfamily) font only within this environment.
  \lstset{basicstyle=\footnotesize\ttfamily}

  \begin{threeparttable}
    \renewcommand{\arraystretch}{1.2}
    \begin{tabularx}{\textwidth}{>{\raggedright\arraybackslash}p{1.5cm}c>{\raggedright\arraybackslash}X>{\raggedright\arraybackslash}l}
      \toprule
      \textbf{Simulator} & \textbf{Issue \#} & \textbf{Issue Description} & \textbf{Impact} \\
      \midrule
      QSim & 565 & \code{QSimSimulator.simulate_expectation_values} for identity operators is wrong. & Incorrect results \\
      Qulacs & 632 & Using \code{ParametricQuantumCircuit} with \code{QuantumCircuitOptimizer} (QCO) causes \code{backprop} to return an empty gradient list when multiple parametric gates are used. & Silent failure \\
      PennyLane & 1086 & Automatic qubit management on \code{lightning.qubit} gives wrong results for \code{probs()}. & Incorrect results \\
      Qiskit Aer & 98 & Incorrect output for CH simulator for some simple circuits with T gates. & Incorrect results \\
      Qiskit Aer & 1351 & Segfault running on \code{AerSimulator} with empty circuit & Crash \\
      Qrack & 234 & 31 qubits or more crash \code{benchmark.cpp} with layers \code{qengine} and \code{qfusion}. & Execution stall \\
      \bottomrule
    \end{tabularx}
    \caption{Example simulator failures in popular quantum simulators. The examples span the reliability surface: logical correctness violations (QSim \#565, Qulacs \#632, PennyLane \#1086, Qiskit Aer \#98), boundary-condition and crash behavior (Qiskit Aer \#1351), and scaling-related execution stalls (Qrack \#234).}
    \label{tbl:simulator-issues}
  \end{threeparttable}
  \vspace{-30pt}
\end{table}

Table~\ref{tbl:simulator-issues} illustrates failures observed in widely used quantum simulators, highlighting the diversity and subtlety of their reliability issues.
Several failures violate fundamental correctness properties without producing explicit errors. For example, QSim~\cite{qsim}~(\#565) computes incorrect expectation values even for identity operators, and Qulacs~\cite{qulacs}~(\#632) silently returns empty gradients during circuit optimization, allowing training to proceed with invalid results. Similar logical correctness violations appear in PennyLane~Lightning~\cite{pennylane-lightning} (\#1086) and Qiskit~Aer~\cite{qiskit-aer}~(\#98), where seemingly simple circuits yield incorrect outputs. Other failures manifest more visibly but still reflect deep execution flaws. Qiskit~Aer~(\#1351) crashes when simulating an empty circuit, exposing boundary-condition errors in simulator initialization, while Qrack~\cite{qrack}(\#234) exhibits execution stalls when scaling beyond 31 qubits, revealing hard limits in execution and resource management.
Together, these examples span logical correctness violations, boundary-condition failures, and scaling-related execution breakdowns, illustrating the reliability surface explored in this study.

Recent empirical work has shown that reliability challenges in quantum software are both widespread and domain-specific. Paltenghi and Pradel~\cite{Paltenghi22} examined 223 bugs across the quantum software stack and found that roughly 40\% are quantum-specific, indicating that many failures stem from concepts unique to quantum computation. Complementing this, Upadhyay et al.~\cite{Upadhyay25} analyzed 157K issues across quantum software repositories and reported that 34\% of issues involve quantum-specific concerns, underscoring the need for quantum-aware software engineering techniques.
Earlier studies by El Aoun et al.~\cite{El_aoun21} further show that a substantial portion of developer difficulties arise during quantum program execution, where simulators play a central role.
While these studies provide important ecosystem-level insights, they treat simulators as part of a broader stack rather than as a distinct class of correctness-critical execution engines, leaving fundamental questions unanswered about how and why quantum simulators fail in practice.

\stitle{Overview of This Work.}
This paper presents a large-scale empirical study of failures in software-based quantum simulators, a class of correctness-critical execution systems that underpin modern quantum programming frameworks. Our goal is to understand how and why simulators fail in practice, which failure modes dominate, and what these failures reveal about the reliability of current simulation infrastructure. To this end, we analyze 394 confirmed issues drawn from 12 widely used open-source quantum simulators, each linked to a concrete fix.
We focus exclusively on classical software simulators that execute quantum programs on classical hardware, and do not consider failures arising from physical quantum devices. Our analysis targets failures that occur during program execution, spanning both quantum-specific execution logic and the surrounding classical software responsible for orchestration, memory management, parameter binding, and cross-language integration.

\begin{figure}[htbp]
  \centering
  \vspace{-10pt}
  \includegraphics[width=\columnwidth]{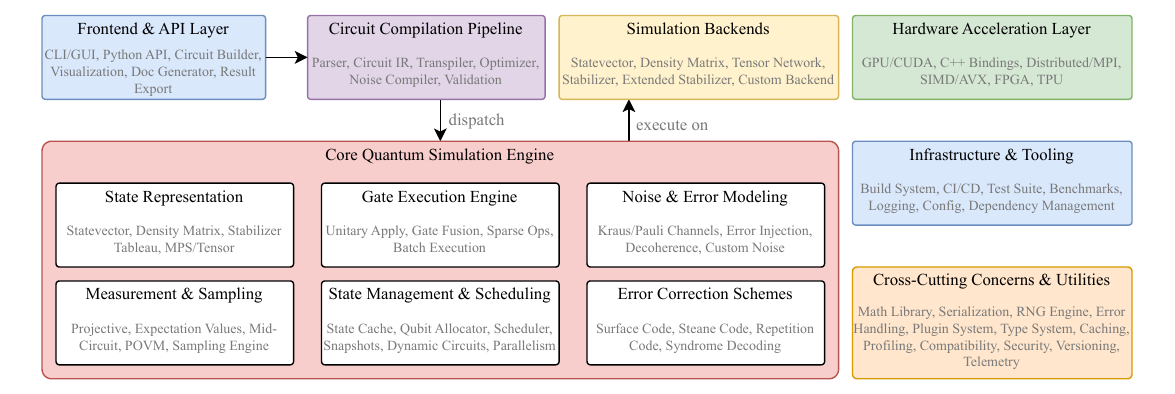}
  \vspace{-20pt}
  \caption{Typical architecture of quantum simulators, showing user-facing interfaces, circuit preparation, simulation backend, and supporting classical layers for acceleration and infrastructure.}
  \label{fig:simulator-arch}
  \vspace{-10pt}
\end{figure}

Figure~\ref{fig:simulator-arch} illustrates the layered architecture of modern software-based quantum simulators.
Across the analyzed repositories, these architectural layers are consistently reflected in repository structure and module organization.
User-facing APIs construct quantum circuits, which are then processed by a compilation pipeline and dispatched to simulation backends such as statevector, density-matrix, tensor-network, or stabilizer engines. These backends rely on a shared simulation core that implements state evolution, gate execution, measurement and sampling, noise modeling, and state management. This core operates within additional classical layers that provide hardware acceleration and infrastructure support, including build systems, testing, configuration, and dependency management. Consequently, simulator correctness depends on the coordinated behavior of both quantum execution logic and extensive classical infrastructure, creating a broad failure surface that affects correctness, performance, and usability.

As illustrated by this architecture, simulator failures can arise from diverse sources and manifest in different ways, ranging from logical correctness violations to crashes, performance stalls, and silent misbehavior.
Understanding these failures, therefore, requires a systematic investigation that cuts across implementation layers, failure manifestations, and development practices. To structure this investigation, we organize our study around the following research questions:

\begin{itemize}[left=0pt]
  \item \emph{RQ1: What are the root causes of issues in quantum simulators?} We investigate the underlying causes of defects to understand whether failures stem from algorithmic errors, numerical instabilities, architectural constraints, or other fundamental sources.
  \item \emph{RQ2: How do issues manifest in quantum simulators?} We examine the observable symptoms and failure modes, such as incorrect results, crashes, performance degradation, or silent errors, to characterize how problems present themselves to users and developers.
  \item \emph{RQ3: Where do critical failures originate in quantum simulators?} We analyze the distribution of failures across different parts of the system, including quantum execution components (e.g., state evolution, gate operations, and measurement) and classical infrastructure (e.g., parameter binding, indexing, memory management, and cross-language interfaces), to identify which layers and components are most prone to critical failures.
  \item \emph{RQ4: How are issues discovered, and what does this reveal about testing effectiveness?} We examine how issues are found through user reports, automated testing, code inspection, comparison testing, and static analysis to understand the effectiveness of current quality assurance practices and identify gaps in testing coverage that allow severe bugs to reach production.
\end{itemize}
These research questions are designed to provide a comprehensive understanding of quantum simulator failures from multiple perspectives. We begin with RQ1 and RQ2, which together characterize simulator issues by examining their underlying root causes and observable manifestations. Understanding both why failures occur and how they present to users yields a clearer picture of the reliability challenges faced by software-based quantum simulators.
To answer these questions, we collect issues and their corresponding fixes from the issue trackers and pull requests of 12 widely used open-source simulators, and focus on 394 confirmed issues that were resolved by accepted patches. Each issue is manually analyzed and labeled with its root cause, its runtime manifestation, and relevant contextual factors such as the affected execution domain and dependencies. Building on this foundation, RQ3 examines where critical failures originate by distinguishing between defects in quantum execution logic (e.g., state evolution, measurement) and defects in surrounding classical infrastructure (e.g., orchestration code, memory management).
Finally, RQ4 investigates how issues are discovered in practice by categorizing discovery mechanisms such as user reports and automated testing, and relating these mechanisms to the types and severities of failures they uncover.
These analyses provide an empirical basis for understanding the reliability surface of quantum simulators. They also reveal gaps in current testing and validation practices that allow critical failures to persist or escape detection.

\stitle{Contributions.}
The key contributions of this work are as follows:

\begin{itemize}[left=0pt]
  \item We construct and analyze a manually validated dataset of 394 resolved issues from 12 widely used open-source quantum simulators, with each issue linked to an accepted fix.

  \item  We provide a systematic analysis of simulator failures along multiple dimensions, including root causes, runtime manifestations, failure origin, and discovery mechanisms.

  \item We derive taxonomies of root causes and manifestations tailored to quantum simulators, enabling consistent comparison across projects and surfacing recurrent failure patterns that are difficult to see from individual repositories.

  \item We identify dominant reliability risks observed in our dataset of simulators, including the prevalence of logical correctness failures, the role of classical orchestration code in triggering severe failures, and the heavy dependence on user reports to uncover critical issues.
\end{itemize}

\noindent\textbf{Key Findings.}
Some representative findings include:

\begin{enumerate}[left=0pt]
  \item Issue discovery is largely user-driven.
    Users report 309/394 issues (78.4\%), while automated testing identifies only 42 (10.7\%). Comparison testing finds 4 issues (1\%), and static analysis identifies only 1 issue.

  \item Severe failures frequently escape internal testing.
    High-impact failures, including crashes, resource exhaustion, and environment-dependent errors, are often not detected during development and are identified after deployment.

  \item Logical correctness failures are common and often silent. Many defects produce incorrect or misleading outputs without triggering explicit errors, making them difficult to detect and particularly risky when simulator outputs are treated as ground truth.

  \item Many critical failures originate outside core quantum execution logic.  A large portion of failures originates from configuration, dependency compatibility, memory management, and other classical systems concerns, rather than from quantum-specific logic.

  \item Ecosystem fragility remains a major source of failures.
    Dependency, packaging, and platform compatibility issues regularly break simulator builds and deployments, contributing substantially to reliability problems observed in practice.
\end{enumerate}

\section{Study Methodology}
\label{sec:method}
\begin{figure}[]
  \centering
  \includegraphics[width=0.98\columnwidth]{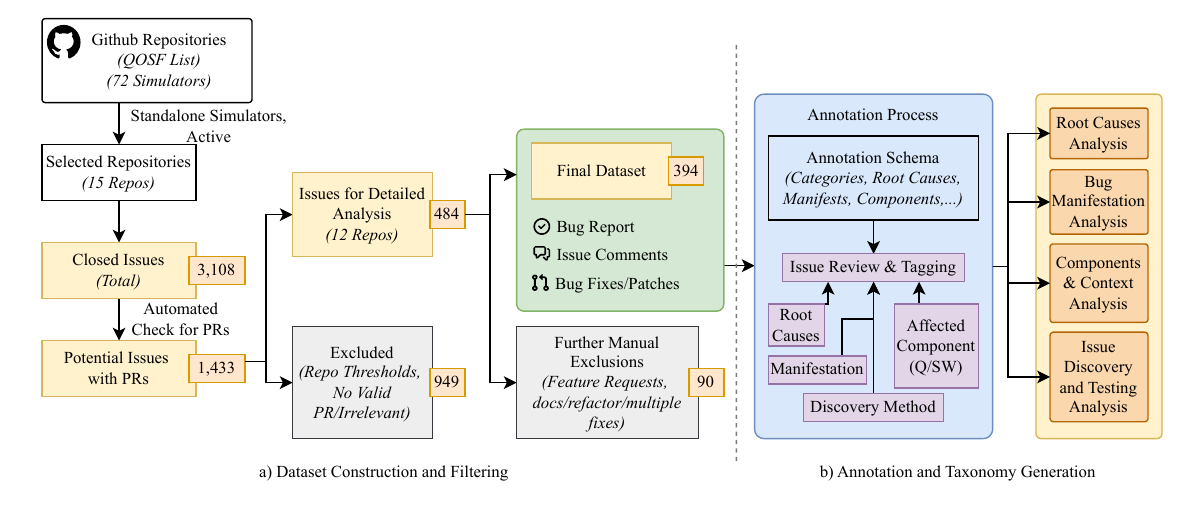}
  \vspace{-20pt}
  \caption{Overview of the data collection and analysis methodology, illustrating the multi-stage pipeline from initial project selection to the final taxonomic analysis of issues.}
  \label{fig:methodology}
  \vspace{-15pt}
\end{figure}

Our data collection and analysis methodology is summarized in Figure~\ref{fig:methodology}.
We begin by identifying a set of actively maintained, standalone software-based quantum simulators and collecting all closed issues from their public GitHub repositories (Section~\ref{sec:dataset_construction}).
To focus on confirmed issues, we retain only issues that are linked to merged pull requests, ensuring that each issue corresponds to a concrete fix. We then manually filter out issues that are out of scope, such as feature requests, documentation-only changes, refactorings, or cleanup commits without functional impact(Section~\ref{sec:collection_filtering}).

The resulting dataset consists of 394 confirmed simulator bugs, each represented by an issue report, associated discussion, and the corresponding code changes that resolved the failure. This dataset serves as the input to our analysis phase (Section~\ref{sec:issue_annotation}).
We iteratively analyze each bug using a structured annotation process that captures its root cause, observable manifestation, affected components, and discovery mechanism. To reduce subjectivity, each issue is independently labeled by two authors across all annotation dimensions.
Disagreements are resolved through discussion until consensus is reached.

\subsection{Dataset Construction}
\label{sec:dataset_construction}
\stitle{Initial Project Identification.}
We began by identifying candidate quantum simulators using the curated list of open-source quantum computing projects maintained by the \gls*{qosf}~\cite{qosf-repo}. This repository provides broad coverage of actively maintained quantum software and is widely used within the community. From this list, we manually reviewed project documentation and repository descriptions to identify projects that provide a functional software-based simulator and are used for executing quantum circuits in practice.

\stitle{Project Selection Criteria.}
Our study focuses on standalone quantum simulators, defined as projects whose primary purpose is to simulate the execution of quantum circuits on classical hardware. We exclude full-featured quantum SDKs that bundle simulators together with compilers, transpilers, hardware interfaces, or cloud services, as failures in such systems are often confounded by non-simulation components.

For widely used quantum SDKs such as IBM Qiskit~\cite{qiskit}, Google Cirq~\cite{cirq}, and PennyLane~\cite{pennylane}, we therefore selected their dedicated simulator backends rather than the parent SDK repositories. Examples include Qiskit Aer, Google’s QSim and Stim simulators, and PennyLane Lightning. These simulators are implemented and maintained as separate repositories with independent issue trackers, allowing us to isolate simulation-specific issues from broader framework concerns such as API design or hardware integration.

\stitle{Project Exclusions.}
We further excluded projects that do not target the execution of quantum programs as they would be run on physical quantum hardware. For example, Quplexity~\cite{quplexity} was excluded because its low-level assembly-oriented design does not align with the circuit-based execution model used by current quantum hardware. Similarly, projects such as Q.js~\cite{qjs}, which primarily support visualization and interactive exploration rather than circuit execution, were excluded. We also removed archived or unmaintained projects, such as HiQSimulator~\cite{hiqsimulator}, to ensure that our dataset reflects contemporary simulator development practices. After applying these criteria, we obtained a set of 15 candidate simulator repositories for further analysis.

\subsection{Issue Collection and Filtering}
\label{sec:collection_filtering}
\stitle{Initial Issue Retrieval.}
For the 15 candidate simulator repositories identified in Section~\ref{sec:dataset_construction}, we collected all closed GitHub issues, yielding a total of 3,108 issues. Since we focus on both failures and their resolutions, we restricted attention to issues that were linked to a pull request (PR) or explicitly referenced a PR in the issue discussion.
Using the GitHub API, we extracted PR identifiers by retrieving both closing PRs associated with each issue and PRs mentioned in issue comments. This automated step produced 1,433 issues with potential PR links.

\stitle{Manual Validation and Preliminary Filtering.}
We then manually reviewed each candidate issue to validate that the extracted PR was genuinely related to the reported problem. During this pass, we performed an initial relevance check and removed issues that were clearly out of scope, including feature requests, documentation updates, and general maintenance tasks. We also excluded issues whose fixes spanned multiple unrelated problems, as such cases complicate attributing a specific fix to a single defect. Figure~\ref{fig:aer128} illustrates an example of a feature request that was excluded at this stage.

\begin{figure}[h]
  \footnotesize
  \begin{example}{Qiskit Aer \#128: Add multi-controlled SWAP gate}\footnotesize
    Add support for a N+2 qubit multi-controlled swap gate.
    The qobj for this gate should be \code{{"name": "mcswap", "qubits": [c1,...,cN, t1, t2]}} where c1,...,c2 are the N control qubits and t1,t2 are the two target qubits that have a SWAP matrix applied to them if all control qubits are in the 1 state.
  \end{example}
  \vspace{-10pt}
  \caption{An example feature request, which requests support for a multi-controlled SWAP gate.}
  \Description{A feature request example showing the specification for implementing an N+2 qubit multi-controlled swap gate in Qiskit Aer, including the required qobj format with control and target qubits}
  \label{fig:aer128}
  \vspace{-10pt}
\end{figure}

\stitle{Repository Threshold Application.}
To ensure sufficient data for meaningful per-project analysis, we required each repository to contain at least five issues with validated PR links.
Applying this threshold led to the exclusion of three repositories: Intel QS~\cite{intel-qs}, Selene~\cite{selene}, and NVIDIA cuQuantum~\cite{cuQuantum}.
Although cuQuantum had a substantial number of reported issues, none were linked to PRs, making it infeasible to analyze how reported problems were resolved.
After applying this criterion, our dataset includes 12 repositories. From the original 3,108 issues, 2,926 remain, among which 484 are linked to candidate fixes.

\stitle{Detailed Manual Analysis.}
We next conducted a detailed manual review of the remaining 484 issues. For each issue, we examined the associated PR to confirm that it contained a concrete fix addressing the reported defect rather than unrelated changes or partial workarounds. We further refined the dataset by excluding issues related to feature additions without underlying bugs, documentation-only changes, refactorings without functional impact, dependency updates that did not resolve specific defects, and general code cleanup. This manual review resulted in the exclusion of 90 additional issues.

\stitle{Final Dataset.}
After completing all filtering stages, our final dataset consists of 394 confirmed simulator issues across 12 repositories, as summarized in Table~\ref{tbl:repo_issue_info}. The selected simulators span a diverse range of simulation techniques, including statevector, stabilizer, tensor-network, and GPU-accelerated backends. This diversity enables us to analyze simulator failures across different execution models and implementation strategies while maintaining a consistent focus on software-based quantum simulation.

\begin{table}[]
  \centering
  \footnotesize
  \caption{Summary of repositories and issues in the study, including total issues, issues linked to pull requests, and selected after manual review.}
  \vspace{-12pt}
  \label{tbl:repo_issue_info}
  \begin{tabular}{lrrrr}
    \toprule
    & \multicolumn{2}{c}{All Issues} & \multicolumn{2}{c}{Selected with Pull Requests~(PRs)} \\
    \cmidrule(lr){2-3} \cmidrule(lr){4-5}
    Repository &   Total &  w/ PRs &     Selected after Analysis &  Excluded after Analysis \\
    \midrule
    \href{https://github.com/pasqal-io/emulators}{Pasqal Emulators}~\cite{pasqal-emulators} &     46 &    9 &      9 &    0 \\
    \href{https://github.com/PennyLaneAI/pennylane-lightning}{PennyLane Lightning}~\cite{pennylane-lightning} &    115 &    32 &      29 &    3 \\
    \href{https://github.com/pasqal-io/pyqtorch}{PyQTorch}~\cite{pyqtorch} &    127 &   16 &      10 &    6 \\
    \href{https://github.com/Qiskit/qiskit-aer}{Qiskit Aer}~\cite{qiskit-aer} &    914 &    223 &       176 &   47 \\
    \href{https://github.com/s-mandra/qflex}{QFlex}~\cite{qflex} &    119 &   19 &       14 &   5 \\
    \href{https://github.com/unitaryfoundation/qrack}{Qrack}~\cite{qrack} &     157 &    17 &       13 &   4 \\
    \href{https://github.com/quantumlib/qsim}{QSim}~\cite{qsim} &     210 &    55 &       48 &   7 \\
    \href{https://github.com/QuEST-Kit/QuEST}{QuEST}~\cite{quest} &     252 &    13 &      12 &    1 \\
    \href{https://github.com/jcmgray/quimb}{QUIMB}~\cite{quimb} &    200 &    6 &      6 &    0 \\
    \href{https://github.com/qulacs/qulacs}{Qulacs}~\cite{qulacs} &     229 &    38 &      33 &    5 \\
    \href{https://github.com/quil-lang/qvm}{QVM}~\cite{qvm} &    137 &   22 &      17 &    5 \\
    \href{https://github.com/quantumlib/Stim}{Stim}~\cite{stim} &     420 &    34 &      27 &    7 \\
    \midrule
    Total &   2,926 &  484 &     394 &  90 \\
    \bottomrule
  \end{tabular}
  \vspace{-15pt}
\end{table}

\subsection{Analyzing Bugs}
\label{sec:issue_annotation}
To analyze the collected simulator bugs, we follow a structured, iterative annotation process inspired by the prior bug analyzing approaches~\cite{chaliasos2021well,xiong2023an,Drosos2024When}. Our goal is to characterize each bug along multiple orthogonal dimensions that capture \myNum{1}~why the bug occurred, \myNum{2}~how it manifested, \myNum{3}~where it originated in the simulator architecture, and \myNum{4}~how it was discovered in practice.

\begin{table}[ht]
  \centering
  \footnotesize
  \vspace{-10pt}
  \caption{Annotation schema and examples.}
  \vspace{-10pt}
  \label{tbl:annotation-detail}
  \renewcommand{\arraystretch}{1.2}
  \lstset{basicstyle=\footnotesize\ttfamily}
  \begin{tabularx}{\textwidth}{p{1.5cm}>{\raggedright\arraybackslash}X>{\raggedright\arraybackslash}X}
    \toprule
    \textbf{Annotation} & \textbf{Description} & \textbf{Examples} \\
    \midrule
    \code{category} & High-level categorization of the issue type & \code{functional-bug}, \code{build-infra} \\
    \code{root} & The fundamental reason the bug exists & \code{algorithmic-flaw}, \code{qubit-truncation} \\
    \code{manifest} & How the bug presents itself to users & \code{wrong-output}, \code{segfault} \\
    \code{discovery} & How the bug was discovered & \code{user-report}, \code{unit-test} \\
    \bottomrule
  \end{tabularx}
  \vspace{-10pt}
\end{table}
\stitle{Annotation Dimensions.}
Each bug in the final dataset is annotated along six dimensions, summarized in Table~\ref{tbl:annotation-detail}.
First, we assign a category label that captures the high-level nature of the issue, such as functional bugs or build and infrastructure failures. Second, we annotate the root cause, describing the fundamental reason the bug exists, for example, algorithmic flaws, incorrect state handling, or dependency-related issues.
Third, we record the manifestation, which captures how the bug presents itself to users, such as incorrect outputs, crashes, execution stalls, or silent failures.

To localize bugs within the simulator architecture, we further annotate two complementary component dimensions.
The quantum component identifies where the bug arises within quantum execution logic, such as state representation, measurement, or sampling.
The software component captures the broader classical subsystem involved, including serialization, backend interfaces, or infrastructure code.
Finally, we annotate the discovery method, indicating how the bug was detected in practice, such as through user reports or automated tests.

\stitle{Annotation Process.}
We perform annotation through an iterative manual review of each issue and its associated pull request, following prior work~\cite{Liu2025}. For each bug, annotators examine the issue description, discussion threads, code changes to fix, and available test cases or reproduction steps. We develop a shared annotation schema through pilot analysis on an initial subset of issues and refine it iteratively as new patterns emerge.

To reduce subjectivity, each bug is independently annotated by at least two authors across all dimensions. During the initial calibration phase (using 100 issues), we assess inter-annotator agreement using Cohen's kappa~\cite{kappa}, obtaining 0.88 for failure manifestations and 0.78 for issue categorization, indicating substantial agreement.
Disagreements are resolved through discussion based on issue reports and associated code changes. If consensus cannot be reached, a third author acts as a tie breaker. This process improves labeling consistency, particularly for cases where failures span multiple components or exhibit indirect symptoms.

\begin{figure}[htbp]
  \footnotesize
  \begin{example}{Qiskit Aer \#416: Segmentation Fault on Pulse Simulator}
    \begin{minipage}[t]{0.40\textwidth}
      \footnotesize
      On certain runs of the \code{pulse_simulator}, a \code{Segmentation fault: 11} immediately halts the program.
    \end{minipage}
    \vspace{-5pt}
    \hfill
    \begin{minipage}[t]{0.56\textwidth}
      \vspace{-1.2\baselineskip}%
       \begin{minted}[fontsize=\scriptsize]{diff}
     if (out != 0):
-        num_times = fc_array.shape[0]
+        num_times = fc_array.shape[0] // 3
         for kk in range(num_times):
        \end{minted}
    \end{minipage}
    \vspace{-5pt}
  \end{example}
  \vspace{-10pt}
  \caption{Qiskit Aer issue \#416 shows error where incorrect loop bounds led to array boundary violations.}
  \label{fig:aer416}
  \vspace{-10pt}
\end{figure}
\stitle{Annotation Example.}

To illustrate our annotation approach, consider issue \#416~\cite{QiskitAerIssue416} from Qiskit Aer as shown in Figure~\ref{fig:aer416}, which involved a segmentation fault caused by an indexing error. The issue was annotated with the following key tags: \code{category:functional-bug}, \code{root:indexing-error}, \code{manifest:segfault}, \code{manifest:crash}, and \code{swcomp:memory-management}. The underlying problem stemmed from incorrect loop bounds calculation in the C/Cython code handling frame changes, where the code used \code{fc_array.shape[0]} instead of \code{fc_array.shape[0] // 3}, leading to array boundary violations and subsequent memory corruption.
\section{RQ1: What are the root causes of issues in quantum simulators?}
\label{sec:rq1}

Understanding why quantum simulators fail is essential for improving the reliability of quantum software tooling.
Based on a detailed analysis of 394 confirmed simulator issues, we identify a diverse set of root causes that span both quantum-specific execution logic and the surrounding classical software infrastructure.
Rather than being confined to a single layer, simulator failures arise from interactions across semantic correctness, numerical computation, resource management, and ecosystem integration.

Table~\ref{tbl:bug-root-causes-taxonomy} summarizes the taxonomy of root causes identified in our study. As in real-world systems, a single issue may involve multiple root causes; therefore, the total count across categories exceeds the 394 issues in our dataset. Overall, root causes concentrate in two broad regions: \myNum{i}~implementation defects that directly compromise the correctness of quantum execution, and \myNum{ii}~ecosystem- and configuration-related failures that prevent correct or reproducible execution despite otherwise sound simulation logic.

\begin{table}[ht]
  \centering
  \footnotesize
  \caption{Taxonomy of root causes identified across 394 confirmed quantum simulator issues, grouped by high-level category and subcategory, with issue counts per subcategory.}
  \vspace{-10pt}
  \label{tbl:bug-root-causes-taxonomy}
  \setlength{\aboverulesep}{0.3ex}
  \setlength{\belowrulesep}{0.3ex}
  \setlength{\abovetopsep}{0pt}
  \setlength{\belowbottomsep}{0pt}
  \renewcommand{\arraystretch}{1.2}
  \newlength{\catwidth}
  \setlength{\catwidth}{1.8cm}
  \setlength{\tabcolsep}{3pt}
  \begin{tabularx}{\textwidth}{>{\centering\arraybackslash}m{\catwidth}>{\raggedright\arraybackslash}p{3.4cm}>{\raggedright\arraybackslash}X>{\centering\arraybackslash}p{0.7cm}}
    \toprule
    \textbf{Category} & \textbf{Subcategory} & \textbf{Description} & \textbf{Count} \\
    \midrule
    \multirow{5}{\catwidth}{\centering Implementation Defects}
    & Algorithmic \& Logic Errors & Incorrect algorithms, formulas, or logic & 186 \\
    & Memory \& Data Management & Memory allocation, copying, or state initialization & 48 \\
    & Indexing \& Boundary Errors & Off-by-one, wrong index, or out-of-bounds & 31 \\
    & Type System \& Data Handling & Type mismatches, casting, or serialization & 49 \\
    & Numerical Computation & Precision loss, instability, overflow, or unit conversion & 21 \\
    \hline
    Quantum State Representation & -- & Qubit mapping, ordering, or truncation errors & 12 \\
    \hline
    \multirow{2}{\catwidth}{\centering Interface \& Integration}
    & API Design \& Usage & Function binding, API misuse, or naming conflicts & 37 \\
    & Concurrency \& Threading & Race conditions, async misuse, deadlocks, or thread-safety & 7 \\
    \hline
    \multirow{2}{\catwidth}{\centering Validation \& Error Handling}
    & Input Validation & Missing or overly strict input checks & 27 \\
    & Error Handling & Uncaught or improperly propagated exceptions & 15 \\
    \hline
    \multirow{2}{\catwidth}{\centering Configuration \& Setup}
    & System Configuration & Incorrect defaults, config files, or build/CI settings & 80 \\
    & Environment Dependencies & Wrong assumptions about environment, imports, or paths & 13 \\
    \hline
    \multirow{3}{\catwidth}{\centering Compatibility}
    & Dependency Compatibility & Deprecated APIs, breaking changes, or incompatible versions & 72 \\
    & Version Compatibility & Language or compiler version incompatibilities & 32 \\
    & Platform Compatibility & Platform-, hardware-, or architecture-specific issues & 37 \\
    \hline
    \multirow{2}{\catwidth}{\centering Resource Management}
    & Resource Limits & Framework-imposed limits or resource exhaustion & 8 \\
    & Scaling \& Performance & Failures at scale or JIT compilation latency & 20 \\
    \hline
    \multirow{2}{\catwidth}{\centering Missing Functionality}
    & Missing Decompositions & Gate or operator decompositions not implemented & 7 \\
    & Other Missing Implementations & Missing features or unimplemented code paths expected by the user & 51 \\
    \bottomrule
  \end{tabularx}
  \vspace{-15pt}
\end{table}

\subsection{Implementation Defects Dominate Root Causes}
Implementation defects constitute the largest class of root causes. Within this category, algorithmic and logic errors are the most prevalent, accounting for 186 issues. These bugs arise from incorrect mathematical formulations, faulty assumptions about quantum operations, or errors introduced when translating theoretical algorithms into executable simulator code. To understand the nature of these algorithmic errors more precisely, we further classify each one by whether resolving it requires knowledge of quantum computing. A quantum-specific algorithmic error is one where understanding or fixing the bug requires knowledge of quantum mechanics, circuit semantics, or quantum-specific data structures. A classical algorithmic error follows patterns common to general software regardless of domain. Among the 186 algorithmic errors, 101 are quantum-specific, while 85 are classical. Within this category, quantum-specific errors are the majority, reflecting that errors in gate semantics, operator composition, and observable computation are genuinely tied to quantum domain knowledge. However, when viewed across all root causes in this taxonomy, classical categories, including memory management, indexing, type handling, configuration, dependency, and platform compatibility, collectively dominate. We revisit this distinction in Section~\ref{subsec:classic-vs-quantum}, where it provides further evidence that classical infrastructure-related issues constitute a large portion of observed failures.

\begin{figure}[htbp]
  \footnotesize
  \vspace{-10pt}
  \begin{example}{Qulacs \#632: Empty gradients when using \code{QuantumCircuitOptimizer}}
    \begin{minipage}[t]{0.35\textwidth}
      \footnotesize
      Using \code{ParametricQuantumCircuit} with \code{QuantumCircuitOptimizer} results in an empty list of gradients as the output of backprop when more than one parametric gate is used.
    \end{minipage}
    \hfill
    \begin{minipage}[t]{0.62\textwidth}
      \vspace{-1.2\baselineskip}
\begin{minted}[fontsize=\scriptsize]{diff}
     if (can_merge_with_swap_insertion(pos, ind1, swap_level)) {
+        if (circuit->gate_list[pos]->is_parametric() ||
+            gate->is_parametric())
+            continue;
         auto merged_gate = gate::merge(circuit->gate_list[pos], gate);
\end{minted}
    \end{minipage}
  \end{example}
  \vspace{-10pt}
  \caption{Qulacs issue \#632 demonstrates a core algorithmic flaw where the optimizer incorrectly merged parametric gates.}
  \label{fig:qulacs632}
  \vspace{-10pt}
\end{figure}

For example, Qulacs \#632~\cite{QulacsIssue632}~(Implementation Defects $\rightarrow$ Algorithmic \& Logic Errors) presented in Figure~\ref{fig:qulacs632} demonstrates how the circuit optimizer incorrectly merged parametric gates into dense matrix representations, causing parameter definitions required for gradient computation to be lost. The resulting silent failure allowed optimization to proceed with empty gradients until explicitly detected.
Similarly, QSim \#576~\cite{QsimIssue576}~(Implementation Defects $\rightarrow$ Algorithmic \& Logic Errors) exposed an incorrect expectation value formula in the C++ core, where the coefficient weight was omitted when evaluating the identity operator, producing systematically wrong results without triggering runtime errors.

\subsection{Memory, Indexing, and Numerical Errors Amplify Failure Severity}
A substantial number of implementation defects stem from memory and data management issues (48 issues) and indexing and boundary errors (31 issues), particularly in simulators that bridge Python frontends with C++ execution. These failures often arise at language and abstraction boundaries, where ownership, lifetime, and mutability assumptions differ, and they tend to amplify the severity of otherwise localized bugs by causing crashes, state corruption, or nondeterministic behavior.

In Qulacs \#74~\cite{QulacsIssue74}~(Implementation Defects $\rightarrow$ Memory \& Data Management), shallow copies introduced by \code{pybind11} bindings caused segmentation faults when Python-managed objects were deallocated while still referenced by C++ code.
A similar boundary-related issue appears in PyQTorch \#35~\cite{PyQtorchIssue35}, where Hamiltonian evolution routines incorrectly mutated the input PyTorch tensor in place, violating caller expectations and corrupting state across successive executions.
These examples illustrate how subtle ownership and mutability mismatches can destabilize simulator execution even when the underlying quantum logic is correct.

\begin{figure}[htbp]
  \vspace{-15pt}
    \begin{minted}[fontsize=\scriptsize]{diff}
    for (UINT i = 0; i < ops.size(); ++i){
-       general_quantum_operator->add_operator(new PauliOperator(ops[i].c_str(), coefs[i]));
+       general_quantum_operator->add_operator(coefs[i], ops[i].c_str());
    }
    \end{minted}
  \vspace{-25pt}
  \caption{Fix for Qulacs \#303 removes dynamic allocation to avoid memory leaks.}
  \label{fig:qulacs303}
  \vspace{-10pt}
\end{figure}

Memory leaks represent another recurring failure mode. In Qulacs \#303~\cite{QulacsIssue303}, newly instantiated \code{PauliOperator} objects were passed into methods that copied their contents but never assumed ownership of the allocated objects, leaving the temporary allocations permanently resident. Figure~\ref{fig:qulacs303} shows how removing dynamic allocation eliminated the leak. Such leaks are particularly harmful in simulators, where long-running experiments and repeated circuit executions can quickly exhaust available memory.

\begin{figure}[htbp]
  \footnotesize
  \vspace{-5pt}
  \begin{example}{Qiskit Aer \#1878: Segfault with specific qubit counts}
    When n equals 39 the circuit has 78 qubits. The simulation step segfaults due to 64 bits integer overflow when number of qubits is so large.
  \end{example}
  \vspace{-10pt}
  \caption{Qiskit Aer issue \#1878 shows integer overflow in memory calculation for large qubit counts.}
  \label{fig:aer1878}
  \vspace{-10pt}
\end{figure}

Numerical computation issues further compound these problems, often producing incorrect results without triggering explicit failures.
Qulacs \#314~\cite{QulacsIssue314} exemplifies a sign error in density-matrix expectation value calculations, where incorrect handling of complex coefficients yielded wrong results whenever observables contained an odd number of \code{Pauli-Y} terms.
Integer overflow presents another common source of failure. In Qrack \#234~\cite{QrackIssue234}, the use of a 32-bit constant (\code{1U}) instead of a 64-bit constant (\code{1ULL}) caused overflow in state-vector indexing for systems exceeding 30 qubits.
A related issue appears in Qiskit Aer \#1878~\cite{QiskitAerIssue1878}, where 64-bit shift operations overflowed in the \code{required\_memory\_mb} function for large qubit counts, producing incorrect memory estimates, as shown in Figure~\ref{fig:aer1878}.

\subsection{Quantum State Representation Errors Are Narrow but Impactful}
Quantum state representation issues arise from incorrect handling of qubit identification, ordering conventions, and allocation decisions. While less frequent than implementation defects, these failures directly affect how a simulator maps circuit-level qubits to its internal state representation, and can therefore trigger severe correctness or resource failures even when the circuit itself is valid. In our dataset, this class accounts for 12 issues.

\begin{figure}[htbp]
  \footnotesize
  \begin{example}{Qiskit Aer \#2249: AerEstimatorV2 fails with circuits transpiled against larger backends}
    Required memory: 18446744073709551615M, max memory: 127903M. The simulator attempted to allocate memory for all 127 qubits in the target backend rather than simulating only the active subset.
  \end{example}
  \vspace{-10pt}
  \caption{Qiskit Aer issue \#2249 demonstrates failure to activate qubit truncation for transpiled circuits.}
  \label{fig:aer2249}
  \vspace{-15pt}
\end{figure}

Qiskit Aer \#2249~\cite{QiskitAerIssue2249}~(Figure~\ref{fig:aer2249}) exemplifies a qubit truncation failure. \code{EstimatorV2} did not activate Aer's internal qubit truncation mechanism for circuits transpiled against large target backends. As a result, the simulator attempted to allocate memory for all 127 qubits supported by the target backend rather than simulating only the active subset used by the circuit, which in turn led to integer overflow in memory calculations and an infeasible memory request.

Qiskit Aer \#997~\cite{QiskitAerIssue997} illustrates a different failure mode, where the MPS backend violated a qubit ordering convention expected by the QASM controller. The backend failed to reset qubit ordering to the sorted convention, causing downstream components to interpret qubit indices inconsistently and resulting in incorrect simulator behavior.

These cases show that state-representation correctness depends not only on implementing quantum operations correctly, but also on enforcing consistent internal conventions for qubit indexing, ordering, and truncation across compilation and execution paths.

\subsection{Ecosystem and Configuration Failures}
Beyond core execution logic, a significant portion of failures originate from the ecosystem and configuration layer. System configuration issues (80 issues) and dependency compatibility problems (72 issues) together form one of the largest clusters in Table~\ref{tbl:bug-root-causes-taxonomy}. These failures include incorrect default settings, CI misconfigurations, and breaking changes in upstream libraries or runtimes.

Across multiple simulators, incompatible Python or NumPy versions, missing compiler flags, or ABI mismatches caused runtime crashes or non-reproducible behavior, even when the underlying simulation logic was correct. These results indicate that simulator reliability depends critically on external infrastructure, not solely on internal correctness.

\subsection{Gaps in Simulator Functionality and Resource Management}
Finally, missing functionality (58 issues) and resource management issues (27 issues) expose gaps between simulator capabilities and user expectations. Missing gate decompositions or unimplemented code paths often prevent otherwise valid circuits from executing. For example, Qiskit Aer \#1447~\cite{QiskitAerIssue1447} (Missing Functionality $\rightarrow$ Missing Decompositions), shown in Figure~\ref{fig:aer1447}, illustrates this class of failures, where the absence of required decomposition rules blocks execution.

\begin{figure}[htbp]
  \footnotesize
  \vspace{-5pt}
  \begin{example}{Qiskit Aer \#1447: RelaxationNoisePass creates noisy circuits that cannot be run}
    RelaxationNoisePass produces noisy circuits that cannot be run directly on Aer simulator when circuits with multi-qubit gates are supplied because the output circuit contains composite instructions labeled like circuit-123.
  \end{example}
  \vspace{-10pt}
  \caption{Qiskit Aer issue \#1447 shows missing decomposition for noise model circuits.}
  \label{fig:aer1447}
  \vspace{-10pt}
\end{figure}

Resource-related failures, including scaling collapse and performance degradation, further demonstrate that correctness cannot be separated from resource-aware design. Code that is logically correct may still fail under realistic workloads due to memory exhaustion or exponential growth in intermediate representations.

\section{RQ2: How do the bugs manifest in quantum simulators?}
\label{sec:rq2}
While RQ1 characterized \emph{why} quantum simulator failures occur, RQ2 examines \emph{how} these failures surface in practice.
We analyze bug manifestations using the taxonomy in Table~\ref{tbl:bug-manifest-taxonomy}, which captures observable failure modes ranging from incorrect results to crashes, resource exhaustion, and degraded performance.
As in real-world systems, a single issue may exhibit multiple manifestations; therefore, the total count across categories exceeds the 394 issues in our dataset.
This perspective reveals not only how users encounter simulator failures but also why many defects are difficult to detect, reproduce, or diagnose.

\begin{table}[]
  \centering
  \footnotesize
  \caption{Bug Manifestation Taxonomy}
  \vspace{-10pt}
  \label{tbl:bug-manifest-taxonomy}
  \renewcommand{\arraystretch}{1.2}
  \newlength{\manwidth}
  \setlength{\manwidth}{2cm}
  \setlength{\tabcolsep}{4pt}
  \begin{tabularx}{\textwidth}{>{\centering\arraybackslash}m{\manwidth}>{\raggedright\arraybackslash}p{3cm}>{\raggedright\arraybackslash}X>{\centering\arraybackslash}p{1cm}}\toprule
    \textbf{Category} & \textbf{Subcategory} & \textbf{Description} & \textbf{Count} \\
    \midrule
    \multirow{2}{\manwidth}{\centering Incorrect Results}
    & Wrong Computation & Produces incorrect numerical or computational results & 104 \\
    & Silent Failures & Fails without any error indication & 15 \\
    \hline
    \multirow{2}{\manwidth}{\centering Data Integrity}
    & Data Corruption & Data or memory becomes corrupted during processing & 12 \\
    & State Inconsistency & Inconsistent behavior across runs or invalid internal state & 34 \\
    \hline
    \multirow{2}{\manwidth}{\centering Process Termination}
    & Crashes & Unexpected termination (crashes or segmentation faults) & 41 \\
    & Hangs \& Blocking & Program becomes unresponsive, hangs, or enters deadlock & 7 \\
    \hline
    \multirow{2}{\manwidth}{\centering Resource Issues}
    & Resource Exhaustion & Runs out of memory or other resources & 11 \\
    & Resource Leaks & Resources not released properly. & 5 \\
    \hline
    Exceptions & - & Unhandled exceptions or assertion failures raised & 136 \\
    \hline
    \multirow{2}{\manwidth}{\centering Startup \& Initialization}
    & Import Failures & Module import fails. & 14 \\
    & Installation \& Startup & Package installation fails or application fails to start & 14 \\
    \hline
    Degraded Performance & - & Performance degradation or slowdowns & 29 \\
    \hline
    Build \& CI Issues & - & Compilation, package build or continuous integration pipeline fails & 76 \\
    \hline
    Test Failures & - & Test suite failures & 34 \\
    \hline
    Warnings & - & Deprecation warnings or user warnings produced & 20 \\
    \bottomrule
  \end{tabularx}
  \vspace{-15pt}
\end{table}

\subsection{Incorrect Results and Silent Failures}

The most common manifestation of simulator defects is incorrect results, comprising 111 issues in total. These include both explicit wrong computations (104 issues) and silent failures (15 issues), where execution completes without errors but produces incorrect or misleading outputs.

\begin{figure}[htbp]
  \footnotesize
  \begin{example}{Qsim \#482: QSimSimulator does not respect \code{invert_mask}}
    \begin{minipage}[t]{0.45\textwidth}
        \begin{minted}[escapeinside=||, fontsize=\scriptsize]{python}
In [3]: |\hl{cirq.Simulator()}|.sample(circuit, repetitions=5)
Out[3]: a
        0  1
        1  1

In [4]: |\hl{qsimcirq.QSimSimulator()}|.sample(circuit, repetitions=5)
Out[4]: a
        0  0
        1  0
    \end{minted}
    \end{minipage}
    \hfill
    \begin{minipage}[t]{0.56\textwidth}
        \begin{minted}[fontsize=\scriptsize]{diff}
+   invert_mask = op.gate.full_invert_mask()
    for j, q in enumerate(meas_indices):
-       results[key][i][j] = full_results[i][q]
+       results[key][i][j] = full_results[i][q] ^ invert_mask[j]
\end{minted}
    \end{minipage}
    \vspace{-15pt}
  \end{example}
  \caption{Qsim issue \#482 shows systematic wrong output due to missing \code{invert_mask} application.}
  \vspace{-10pt}
  \label{fig:qsim482}
\end{figure}

Qsim \#482~\cite{QsimIssue482}~(Figure~\ref{fig:qsim482}) illustrates a systematic wrong-result failure caused by missing application of the \code{invert\_mask} during sampling. Although the circuit executes successfully, the simulator returns incorrect measurement outcomes that violate expected semantics. Because no error is raised, users may incorrectly trust the results.

\begin{figure}[htbp]
  \footnotesize
  \begin{example}{Qiskit Aer \#1479: Multi-chunk parallelization not applied}
    Multi-chunk parallelization is not applied when using \code{density_matrix} method with noise models. Expected behavior: \code{cache_blocking} metadata will be recorded in the result when applied.
  \end{example}
  \vspace{-10pt}
  \caption{Qiskit Aer issue \#1479 demonstrates silent failure where parallelization is disabled.}
  \label{fig:aer1479}
  \vspace{-10pt}
\end{figure}

Silent failures are even more problematic. Qiskit Aer \#1479~\cite{QiskitAerIssue1479} shown in Figure~\ref{fig:aer1479} demonstrates a case where multi-chunk parallelization was silently disabled for density-matrix simulations with noise models. The simulator did not emit warnings or errors; instead, users experienced severe performance degradation without understanding the underlying cause. Such failures undermine both correctness and performance assumptions while remaining invisible to correctness checks.

\subsection{Data Integrity and State Inconsistency}

Data integrity violations occur when internal quantum state representations become corrupted or inconsistent during execution. Table~\ref{tbl:bug-manifest-taxonomy} shows 46 such cases, including state inconsistency across runs and memory corruption.

Qsim \#283~\cite{QsimIssue283} exemplifies a copy-versus-reference error in which \code{QTrajectorySimulator} directly mutated the caller's state vector instead of performing a deep copy. This led to unintended modification of input state and incorrect subsequent computations.

\begin{figure}[htbp]
  \vspace{-5pt}
  \footnotesize
  \begin{example}{Qiskit Aer \#1308: Nondeterministic MPS with multiple threads}
    The MPS simulator gives nondeterministic results when executed with \code{max_parallel_threads} greater than 1 even when seeding the simulator. RNG seeds are not getting set properly across threads.
  \end{example}
  \vspace{-10pt}
  \caption{Qiskit Aer issue \#1308 shows race condition in random number generation across threads.}
  \label{fig:aer1308}
  \vspace{-10pt}
\end{figure}

State inconsistency issues are further illustrated by Qiskit Aer \#1308~\cite{QiskitAerIssue1308}~(Figure~\ref{fig:aer1308}), where the MPS backend produced nondeterministic results when executed with multiple OpenMP threads. Even with a fixed random seed, concurrent access to the shared C++ random number generator caused race conditions, violating determinism guarantees expected by users.

\subsection{Crashes, Hangs, and Resource Failures}
The most severe manifestations of simulator defects involve abnormal process termination or loss of responsiveness, including crashes, segmentation faults, hangs, and unrecoverable runtime errors.
While these failures are immediately visible to users, they typically arise from latent defects that remain dormant until triggered by scale, platform differences, or repeated execution.

\begin{figure}[htbp]
  \footnotesize
  \vspace{-5pt}
  \begin{example}{Qiskit Aer \#755: Segmentation Fault with \code{density_matrix_gpu}}
    After repeated execution of circuits, the density matrix CPU simulator will cause segmentation fault and the GPU simulator will cause \code{cudaErrorInvalidConfiguration}. The error typically happens around 100 to 200 loops.
  \end{example}
  \vspace{-10pt}
  \caption{Qiskit Aer issue \#755 demonstrates crash from missing memory validation checks.}
  \label{fig:aer755}
  \vspace{-10pt}
\end{figure}

Qiskit Aer \#755~\cite{QiskitAerIssue755}~(Figure~\ref{fig:aer755}) exemplifies a crash caused by missing memory validation in the density-matrix backend. As circuit execution repeats, the simulator attempts to allocate state vectors exceeding available resources, resulting in segmentation faults on the CPU and runtime failures on the GPU. This illustrates how unchecked assumptions in memory requirement calculations can remain undetected during testing but surface under sustained execution or larger workloads.

Beyond crashes, several issues manifest as runtime errors or startup failures that prevent simulators from operating correctly. QFlex \#284~\cite{QFlexIssue284} reports a runtime exception caused by incorrect assumptions in the C++ tensor copy routine, which misclassified rank-0 tensors as uninitialized and falsely reported a memory leak. PennyLane-Lightning \#243~\cite{PennyLaneLightningIssue243} demonstrates an import failure triggered by an upstream dependency change that renamed a core module, causing the simulator to fail during initialization. Stim \#373~\cite{StimIssue373}  illustrates a blocking startup failure where platform-specific API usage invoked \code{os.sched_setaffinity} without guarding for its absence on macOS, leading to a hang during simulator initialization.

To summarize, crash-level and blocking failures often arise from implicit assumptions about memory availability, dependency stability, and platform support.
While less frequent than incorrect or silent results, they have a high impact by preventing execution and exposing weaknesses in validation, portability, and stress testing.

\subsection{Build, Test, and Startup Failures}

A significant number of failures occur prior to execution, during build, testing, or initialization. Build and CI failures are observed in 76 issues, test failures in 34, and startup or installation problems in 28. Although not directly tied to quantum execution logic, these failures have a direct impact on usability and deployment.

Build failures commonly arise from fragile assumptions about toolchains and dependencies. Qsim \#625~\cite{QsimIssue625} illustrates this pattern, where an upstream update to LLVM~\cite{llvm} and \code{libc-bin} caused compilation failures without changes to simulator code. Test failures often reflect hidden platform assumptions: PennyLane-Lightning \#209~\cite{PennyLaneLightningIssue209} failed on Windows due to tests referencing the unavailable \code{numpy.complex256} type.

Startup and initialization issues similarly stem from dependency mismatches or incompatible environments. Together, these failures highlight that simulator reliability depends not only on execution correctness but also on robust build, test, and dependency management practices.

\subsection{Performance Degradation and Warnings}
Finally, a smaller but important set of issues manifests as performance degradation (29 issues) or warnings (20 issues). Performance regressions do not break correctness but can reduce usability, especially for simulators supporting large-scale experiments or iterative workflows. These issues often arise from mismatches between implementation choices and underlying frameworks. For example, Pasqal-io Emulators \#54~\cite{PasqalEmulatorsIssue54} performs bitstring sampling in a shot-by-shot loop rather than using batched tensor operations, preventing efficient vectorization and causing slowdowns.

Warnings are less disruptive but signal deeper compatibility or maintenance issues. Although execution proceeds, they often expose technical debt that can later evolve into correctness failures. Qiskit Aer \#2178~\cite{QiskitAerIssue2178} illustrates this pattern, where a deprecation warning due to inheritance from \code{ProviderV1} reveals reliance on outdated APIs that may break as the ecosystem evolves.

\section{RQ3: Where do critical failures originate in quantum simulators?}
\label{sec:rq3}

Algorithmic and logic defects constitute the largest root-cause category in our dataset, accounting for 186 of the 394 confirmed issues.
Unlike memory errors or configuration problems, these defects directly affect the semantic correctness of quantum circuit execution. To better understand their nature, we perform a deeper analysis in two stages. First, we distinguish between classical algorithmic errors and quantum-specific algorithmic errors. Second, we analyze the distribution of quantum-specific errors across simulator components to identify structural concentration points.

\subsection{Classical vs Quantum-Specific Algorithmic Failures}
\label{subsec:classic-vs-quantum}
To better understand the structure of algorithmic and logic defects, we further distinguish whether resolving each issue requires quantum domain knowledge.
This refinement separates errors that arise from general programming logic from those that stem directly from quantum semantics.

A \emph{quantum-specific algorithmic error} requires understanding quantum mechanics, circuit semantics, or quantum-specific data structures.
These defects involve violations of mathematical invariants governing state evolution, operator composition, or observable evaluation. For example, Qsim \#576~\cite{QsimIssue576} required recognizing that identity operator contributions must retain their coefficient weights when computing expectation values.
Similarly, Qulacs \#314~\cite{QulacsIssue314} depended on understanding that Pauli-Y operators introduce sign changes in density matrix calculations due to their imaginary eigenvalues. In both cases, diagnosing the issue required reasoning about quantum linear algebra rather than conventional software patterns.

In contrast, a \emph{classical algorithmic error} follows patterns common to general systems software. Stim \#926~\cite{StimIssue926} illustrates this distinction: the implementation failed to handle negative Python indices when converting to C++ \code{size_t}, resulting in incorrect boundary behavior. This defect reflects a standard type-conversion oversight and requires no quantum-specific expertise to identify.

Of the 186 algorithmic defects, 101 (54.3\%) are quantum-specific, and 85 (45.7\%) are classical. Thus, quantum semantics account for a slight majority within this root-cause category. These quantum-specific defects primarily involve gate semantics, operator composition, measurement handling, and observable computation.

However, when considered alongside the other root causes identified in RQ1 (Sec.~\ref{sec:rq1}), the overall failure landscape remains predominantly classical.
Categories such as configuration errors, dependency incompatibilities, memory management issues, indexing errors, and platform mismatches collectively account for a larger portion of the dataset than quantum-specific algorithmic defects. This contrast highlights that while quantum semantics introduce distinct correctness challenges, simulator reliability is equally shaped by conventional infrastructure-level engineering concerns.

\begin{wrapfigure}{r}{0.44\textwidth}
  \vspace{-10pt}
  \footnotesize
  \centering
  \includegraphics[width=0.44\columnwidth]
  {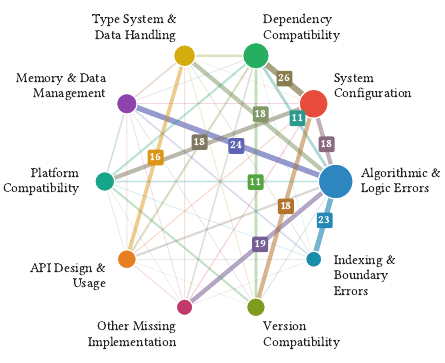}
  \vspace{-25pt}
  \caption{Co-occurrence network showing correlations between root cause categories.}
  \vspace{-10pt}
  \label{fig:category-correlation-network}
\end{wrapfigure}

The co-occurrence structure in Figure~\ref{fig:category-correlation-network} further clarifies how these defects interact. Algorithmic and logic errors form a hub, showing strong associations with memory and data management (24 co-occurrences), indexing and boundary errors (23), system configuration (18), and missing implementation (19). This indicates that algorithmic defects rarely occur in isolation; instead, they often co-occur with classical infrastructure failures. Compatibility and configuration issues also form a dense cluster, suggesting that evolving dependencies and environment assumptions systematically influence simulator behavior. The strong coupling among the algorithmic, memory, and indexing categories highlights that quantum-specific logic is tightly intertwined with low-level system concerns, allowing failures to propagate across abstraction boundaries rather than remain localized.

\subsection{Component-Level Concentration of Quantum-Specific Errors}
\label{subsec:rq3-components}
We next analyze the 101 quantum-specific algorithmic errors by simulator component, as shown in Table~\ref{tbl:quantum-algo-components}. The distribution reveals a clear structural concentration.

\begin{table}[]
  \centering
  \footnotesize
  \caption{Distribution of 101 quantum-specific algorithmic errors by quantum component.}
  \vspace{-10pt}
  \label{tbl:quantum-algo-components}
  \renewcommand{\arraystretch}{1.2}
  \setlength{\tabcolsep}{4pt}
  \begin{tabularx}{\textwidth}{>{\raggedright\arraybackslash}p{2.4cm}>{\raggedright\arraybackslash}X>{\centering\arraybackslash}p{1.0cm}}
    \toprule
    \textbf{Component} & \textbf{Description} & \textbf{Count} \\
    \midrule
    State Simulation   & Errors in evolving multi-qubit state vectors or density matrices, including statevector updates, normalization, and backend-specific evolution logic & 20 \\
    Noise Model        & Incorrect application or composition of error channels, superoperators, or decoherence models on quantum gates or circuits & 16 \\
    Gate Operations    & Bugs in gate matrix definitions, phase conventions, controlled gate decompositions, or gate application order & 13 \\
    Circuit Execution  & Errors in sequencing gate operations, handling barriers, managing classical registers, or processing conditional instructions & 12 \\
    MPS                & Failures in matrix product state contraction order, bond dimension management, or canonical form maintenance during evolution & 9 \\
    Measurement        & Incorrect sampling logic, qubit ordering during readout, post-selection semantics, or probability computation & 8 \\
    Gradient / Diff.   & Errors in parameter-shift rules, adjoint differentiation, or backpropagation through parametric quantum gates & 7 \\
    Expectation Values & Incorrect evaluation of quantum observables, including missing coefficient weights or wrong operator composition & 6 \\
    Stabilizer         & Bugs in Pauli frame tracking, stabilizer tableau updates, or Clifford operation application & 6 \\
    Error Correction   & Errors in syndrome decoding logic, logical qubit encoding, or fault-tolerant operation sequences & 3 \\
    Quantum Algorithms & Incorrect implementation of a specific quantum algorithm routine & 1 \\
    \midrule
    Total & & 101 \\
    \bottomrule
  \end{tabularx}
  \vspace{-10pt}
\end{table}

State simulation, noise modeling, gate operations, and circuit execution together account for 61 of the 101 quantum-specific defects (60.4\%). These components are responsible for evolving multi-qubit state representations, composing operators, and coordinating classical–quantum execution flow. Errors in these layers typically violate invariants such as correct parameter binding, operator composition, qubit ordering, or state dimension consistency.

\begin{figure}[htbp]
  \footnotesize
  \begin{example}{Qiskit Aer \#1849: Parameter binding errors after conditional instruction injection}
    \begin{minipage}[t]{0.40\textwidth}
      Internal instruction injection for conditionals shifts C++ instruction indices, invalidating Python-side parameter binding.
    \end{minipage}
    \hfill
    \begin{minipage}[t]{0.56\textwidth}
      \vspace{-1.2\baselineskip}
    \begin{minted}[fontsize=\scriptsize]{diff}
def assemble_circuit(circuit: QuantumCircuit):
+    num_of_aer_ops = 0
+    index_map = []
     for inst in circuit.data:
         if inst.operation.condition:
             aer_circ.bfunc(...)
+            num_of_aer_ops += 1
+        num_of_aer_ops += _assemble_op(...)
+        index_map.append(num_of_aer_ops - 1)
-    return aer_circ
+    return aer_circ, index_map
    \end{minted}
    \end{minipage}

  \end{example}
  \vspace{-10pt}
  \caption{Qiskit Aer issue \#1849 demonstrates parameter binding failure when internal instructions shift indices.}
  \label{fig:aer1849}
\end{figure}

For example, Qiskit Aer \#1849~\cite{QiskitAerIssue1849}~(Figure~\ref{fig:aer1849}) exposed a parameter-binding error introduced after conditional instruction injection. Internal circuit rewriting shifted parameter indices, causing incorrect parameter resolution during execution. The defect did not involve memory corruption or configuration problems; rather, it reflected fragility in maintaining semantic consistency during circuit transformation.

Similarly, Qsim \#327~\cite{QsimIssue327} as shown in Figure~\ref{fig:qsim327} miscomputed state vector size by using the set of ordered qubits instead of all qubits, resulting in incorrect dimension allocation. This issue arose from a subtle mismatch between logical circuit representation and physical state indexing—an invariant central to state simulation correctness.

\begin{figure}[htbp]
  \footnotesize
    \begin{minted}[fontsize=\scriptsize]{diff}
-   qubits = self.all_qubits()
-   qsim_circuit.num_qubits = len(qubits)
    ordered_qubits = cirq.ops.QubitOrder.as_qubit_order(qubit_order).order_for(
-     qubits)
+     self.all_qubits())
+   qsim_circuit.num_qubits = len(ordered_qubits)
    \end{minted}
  \vspace{-20pt}
  \caption{Fix for Qsim \#327 uses ordered qubits instead of all qubits for size calculation.}
  \label{fig:qsim327}
\end{figure}

\begin{figure}[htbp]
  \vspace{-10pt}
  \footnotesize
  \begin{example}{PyQTorch \#330: GPSR with repeated parameter cannot be differentiated twice with shots}
    With the current implementation, second order derivative do not work when asking for shots. The context used in the GPSR differentiation method fails to save the right values for the backward pass.
  \end{example}
  \vspace{-10pt}
  \caption{PyQTorch issue \#330 demonstrates parameter context loss during backpropagation.}
  \label{fig:pyqtorch330}
  \vspace{-10pt}
\end{figure}

Advanced simulation techniques introduce additional sources of complexity. Modules implementing gradient computation and differentiable quantum operations account for 7 issues, while MPS, stabilizer, and expectation-value components contribute another 21 combined. These subsystems depend on non-trivial invariants such as contraction order, canonical form maintenance, and correct propagation of parameter-shift derivatives.
PyQTorch \#330~\cite{PyQtorchIssue330} (Fig.~\ref{fig:pyqtorch330}) illustrates this fragility, where repeated parameters in a circuit are not properly differentiated twice, leading to incorrect gradient computation. The error stems from incomplete handling of parameter reuse in the backpropagation logic, demonstrating how differentiable extensions expand the semantic~error~surface.

Overall, the component distribution shows that quantum-specific algorithmic defects concentrate where high-dimensional linear algebra, circuit rewriting, and classical–quantum coordination intersect. These layers encode complex mathematical invariants that are rarely formally specified or automatically validated, making subtle semantic deviations both likely and difficult to detect.

\section{RQ4: How are issues discovered, and what does this reveal about testing effectiveness?}
\label{sec:rq4}

Next, we examine how simulator issues are discovered in practice and what these discovery patterns reveal about the effectiveness of existing testing and quality assurance mechanisms. Table~\ref{tbl:bug-discovery} summarizes the distribution of discovery methods across all 394 issues, including user reports, automated testing, code inspection, comparison testing, CI runs, and static analysis.

\subsection{User Reports Dominate Issue Discovery}
User reports are by far the dominant discovery mechanism, accounting for 309 issues (78.4\%). This reliance on users is especially pronounced for severe failures. Users discover 89.2\% of exceptions, 81.2\% of segmentation faults, and 84.8\% of crashes. Notably, every memory exhaustion and resource leak issue in the dataset was first reported by users.

These results indicate that many critical failures escape internal testing and are encountered only under real-world usage conditions, such as large workloads, extended execution, or platform-specific configurations. The prevalence of user discovery suggests that quantum simulators rely heavily on their user communities as the final layer of quality assurance. Specifically, 309 of 394 issues were initially reported by users rather than detected internally.

\subsection{Effectiveness and Limits of Automated Testing}
Automated testing, including unit, CI, and integration tests, identifies a limited portion of issues.
As shown in Table~\ref{tbl:bug-discovery}, testing accounts for only a small fraction of bug discovery overall. Unit tests are most effective when failure modes are explicitly anticipated, particularly for incorrect output cases. However, their effectiveness remains constrained by the scope of test inputs and~expected~behaviors.

\begin{table}[t]
  \centering
  \footnotesize
  \caption{Bug discovery mechanisms. Issues are identified through user reports, code review, testing activities, CI runs (including build and execution failures), and static analysis.}
  \vspace{-10pt}
  \label{tbl:bug-discovery}
  \begin{tabular}{lrr}
    \toprule
    \textbf{Discovery Method} & \textbf{Count} & \textbf{Percentage} \\
    \midrule
    User report         & 309 & 78.43\% \\
    Code review         &  26 &  6.60\% \\
    CI pipeline (excluding testing)         & 16 & 4.06\% \\
    \midrule
    Testing  & 42 & 10.66\% \\
    \qquad Unit test           & 24 & 6.09\% \\
    \qquad Integration test    & 14 & 3.55\% \\
    \qquad Comparison testing  &  4 & 1.02\% \\
    \midrule
    Static analysis     &   1 &  0.25\% \\
    \bottomrule
  \end{tabular}
\end{table}
This limitation becomes evident for runtime failures. Unit tests detect only a small set of crashes~(2 of 33) and segmentation faults (4 of 32), and fail to capture memory exhaustion and resource leak issues. These failures typically emerge under large-scale or resource-intensive conditions that are not exercised in typical test scenarios. As a result, automated testing tends to focus on small inputs and controlled environments, leaving scale-dependent and system-level failures largely undetected.

\begin{figure}[htbp]
  \vspace{-5pt}
  \footnotesize
  \begin{example}{Qrack \#681: Q\# \code{SWAPTwoQubits} unit test failure}
    The \code{SWAPTwoQubits} unit test for Q\# fails. If \code{QUnit} is removed from the \code{PInvoke} API in favor of \code{QPager}, the test passes, suggesting that the problem is in \code{QUnit} layer. There might be a bug in \code{ProbParity()} or \code{MParity()}.
  \end{example}
  \vspace{-10pt}
  \caption{Unit test caught layer-specific quantum gate implementation bug.}
  \label{fig:qrack681}
  \vspace{-10pt}
\end{figure}

Concrete examples illustrate both the strengths and limitations of unit testing. Qulacs unit tests detect incorrect gradient computation when parametric gates are improperly merged during optimization, while Stim unit tests expose boundary violations in polygon drawing logic. Similarly, Qrack \#681~\cite{QrackIssue681} (Figure~\ref{fig:qrack681}) identifies a faulty implementation of the Q\# \code{SWAPTwoQubits} operation. These cases show that unit tests are effective at catching well-specified, localized errors, but their effectiveness is limited to narrowly defined scenarios.

\subsection{CI and Integration Testing Focus on Infrastructure}
CI pipelines and integration testing primarily expose issues related to build systems, configuration, and dependency compatibility rather than core execution correctness. These failures often stem from mismatches between software environments, toolchains, or external libraries, reflecting the complexity of maintaining a stable execution environment.

CI pipelines (excluding testing) capture failures during build and execution stages, including compiler incompatibilities, missing dependencies, and deprecated runtime configurations. In our dataset, CI pipelines identify 16 of 29 CI-specific failures and 9 out of 63 build-related issues. These failures reflect environment and infrastructure inconsistencies rather than defects in simulation~logic.

\begin{figure}[htbp]
  \footnotesize
  \vspace{-5pt}
  \begin{example}{Qiskit Aer \#741: Pulse system model using dict based access}
    There is a pulse simulator test failure. After PR 695 the \code{calculate_channel_frequencies} is still using dict based access despite not using a dict anymore. L220 should be updated from \code{if u_lo_idx['q']} to \code{if u_lo_idx.q}.
  \end{example}
  \vspace{-10pt}
  \caption{Qiskit Aer issue \#741 shows integration test detecting API usage mismatch after refactoring.}
  \label{fig:aer741}
  \vspace{-10pt}
\end{figure}

Integration tests reveal broader system-level inconsistencies. They detect 6 of 34 test failures and uncover compatibility issues involving external libraries and APIs. Qiskit Aer \#741~\cite{QiskitAerIssue741}~(Figure~\ref{fig:aer741}) illustrates this role: an integration test exposed outdated dictionary-based access after a refactoring change. While effective for infrastructure validation, these techniques catch only a small fraction of crashes, segmentation faults, or incorrect results.

\subsection{Code Inspection Finds Logical Errors but Misses Runtime Failures}
Code inspection accounts for 26 discovered issues and identifies a diverse range of manifestations, including incorrect results, exceptions, build failures, warnings, and performance degradations. It is particularly effective at uncovering logical errors, incorrect assumptions, and missing validation checks through reasoning about code structure.

In contrast, failures that depend on execution conditions, such as crashes, segmentation faults, hangs, and resource exhaustion, are rarely identified through inspection in our dataset. These failures typically arise from dynamic behavior, including concurrency, memory pressure, and platform-specific interactions, which are not directly visible from code alone.

This distinction highlights the complementary roles of inspection and dynamic testing. Code inspection is well-suited for reasoning about program logic, while execution-based techniques are necessary to expose failures that emerge only under specific runtime conditions.

\subsection{Comparison Testing and Static Analysis Are Rarely Used}
Comparison testing against reference implementations or analytical results is used sparingly, uncovering only four issues in total. Of these, three are wrong output issues, highlighting the potential effectiveness of this approach for numerical correctness.
\begin{figure}[htbp]
  \footnotesize
  \vspace{-5pt}
  \begin{example}{Qiskit Aer \#1193: Incorrect results when using \code{cache_blocking}}
    Running the test teleport circuit on the statevector simulator with cache blocking enabled gives the wrong result.
  \end{example}
  \vspace{-10pt}
  \caption{Qiskit Aer issue \#1193 demonstrates comparison testing revealing optimization correctness bug.}
  \label{fig:aer1193}
  \vspace{-10pt}
\end{figure}

Qiskit Aer \#1193~\cite{QiskitAerIssue1193} demonstrates this pattern, where comparing results with and without cache blocking revealed incorrect behavior. Despite well-defined mathematical properties and the availability of multiple simulator implementations, comparison testing remains largely absent.

Static analysis tools identify only one issue in the entire dataset: an assertion failure detected by \code{AddressSanitizer}~\cite{Konstantin2012AddressSanitizer} during test execution. This absence is striking given the prevalence of memory-related failures, including 32 segmentation faults, 6 memory corruption issues, and 6 data corruption issues, which fall within the strengths of memory safety tools.

\subsection{Implications for Testing Effectiveness}

Taken together, the discovery patterns reveal systematic gaps in current testing practices.
Automated tools are effective at detecting test regressions and infrastructure failures but consistently miss severe bugs, including crashes, segmentation faults, memory exhaustion, incorrect results, and silent failures.
These issues frequently surface only under edge conditions, large workloads, or platform-specific environments that automated tests do not exercise.

The dominance of user-discovered wrong outputs, incorrect state, and silent failures suggests that existing tests do not provide correctness guarantees.
Property-based testing~\cite{MacIver2019Hypothesis} offers a promising direction to address this gap.
Prior work has demonstrated the feasibility of applying property-based testing to quantum programs~\cite{Honarvar2020}, and coverage-based approaches with well-defined test oracles have shown effectiveness in detecting faults that example-based unit tests miss~\cite{Ali2021}.
By generating random circuits and validating invariants such as unitarity preservation, probability normalization, and measurement distribution correctness, such testing could expose subtle mathematical and semantic errors that evade example-based unit tests.
Importantly, these properties are well defined for quantum execution, making them suitable targets for systematic validation.

Memory safety tooling represents another underutilized resource.
Given the prevalence of segmentation faults, memory corruption, and cross-language boundary failures in our dataset, integrating tools such as \code{AddressSanitizer}~\cite{Konstantin2012AddressSanitizer} and \code{ThreadSanitizer}~\cite{Serebryany2009ThreadSanitizer} into CI pipelines would directly target failure modes that automated tests currently miss.
Similarly, differential testing~\cite{MacKeeman1998Differential} across independent simulator backends could systematically expose semantic inconsistencies introduced by optimizations, particularly given that multiple mature simulator implementations exist and quantum operations have well-defined mathematical ground truth.

\section{Implications and Discussion}
\label{sec:implications}
Our study exposes a gap between the perceived reliability of quantum simulators and their behavior in practice.
The findings have practical implications for researchers and developers, enhancing understanding of project dynamics and informing best practices for advancing quantum software.

\subsection{Findings}
\stitle{Finding 1: Quantum simulator reliability is constrained more by classical infrastructure than quantum logic.}
Many failures originate from conventional software issues such as memory management, configuration, dependency compatibility, and indexing errors, rather than from the implementation of quantum algorithms. While quantum-specific defects do occur, they are typically localized within core execution components. In contrast, infrastructure-related issues appear across the entire system stack and account for a substantial portion of failures (Table~\ref{tbl:bug-root-causes-taxonomy}). This imbalance suggests that improving simulator reliability requires advances in classical systems engineering as much as in quantum algorithm design.

\stitle{Finding 2: Silent incorrect results represent a fundamental and under-recognized failure mode.}
A large class of simulator bugs does not manifest as crashes or explicit errors but instead produces plausible yet incorrect outputs. These failures are particularly difficult to detect because execution appears successful, masking underlying correctness violations. As reflected in Table~\ref{tbl:bug-manifest-taxonomy}, incorrect results constitute one of the most common manifestations, including both explicit errors and silent failures. This pattern highlights a key challenge for validation, as traditional failure signals are often absent.

\stitle{Finding 3: Real-world failures are poorly captured by existing testing practices.}
In practice, many severe failures are not detected during development but are instead discovered by users after deployment. These include crashes, resource exhaustion, and environment-dependent failures that escape standard testing pipelines. As illustrated in Table~\ref{tbl:bug-discovery}, automated testing identifies only a limited subset of issues. This gap indicates that current testing approaches do not adequately reflect real-world usage conditions or system complexity.

\stitle{Finding 4: Simulator failures concentrate in quantum operations that require preserving global semantic invariants.}
Quantum-specific defects are not uniformly distributed across components but instead cluster in operations such as state evolution, gate application, circuit execution, and noise modeling (Table~\ref{tbl:quantum-algo-components}). These operations implement core properties of quantum computation, including linearity, unitary evolution, non-commutativity, and probabilistic measurement. Correctness in these settings requires maintaining global invariants, such as state normalization, consistent ordering of operations, and coherent propagation of transformations across the entire~system.

\subsection{Actionable Suggestions and Takeaways}

\stitle{Suggestion 1 (For Simulator Developers): Strengthen systems-level reliability safeguards.}
Given the dominance of classical implementation defects, simulator developers should prioritize memory safety instrumentation, overflow detection, and boundary checking. Integrating tools such as \code{AddressSanitizer} and \code{ThreadSanitizer} into CI pipelines would directly target segmentation faults and memory-related issues observed in our dataset. Stress testing at large qubit scales should be routinely incorporated to expose overflow and allocation defects early.

\stitle{Suggestion 2 (For Simulator Developers): Incorporate invariant-based and property-based testing.}
The high user discovery rate for wrong outputs and silent failures indicates insufficient correctness validation. Property-based testing can address this gap by generating randomized circuits and verifying invariants such as probability normalization, unitary preservation, deterministic seeding behavior, and consistency across repeated executions. Differential testing across multiple simulators can further detect optimization-induced semantic deviations.

\stitle{Suggestion 3 (For Framework and SDK Developers): Stabilize and monitor interface boundaries.}
A substantial portion of failures arise from dependency changes, API misuse, and compatibility mismatches. Framework developers should maintain explicit interface contracts with simulator backends and perform compatibility testing across supported versions. Version pinning strategies and automated dependency audits can mitigate breakages caused by upstream changes.

\stitle{Suggestion 4 (For Tool Builders and Infrastructure Engineers): Expand CI coverage across platforms and resource configurations.}
Platform-specific issues systematically evade detection when CI matrices are limited. Expanding CI to include Windows, macOS, ARM64, and GPU configurations would catch many environment-dependent defects earlier. Continuous performance regression monitoring should also be incorporated to detect degradations before release.

\stitle{Suggestion 5 (For Researchers and Simulator Users): Validate experimental results across multiple execution paths.}
Given the prevalence of silent incorrectness, researchers should avoid relying on a single simulator configuration when validating results. Cross-validation across simulators, execution modes, or optimization settings can reduce the risk of drawing conclusions from incorrect outputs. Explicit verification of numerical invariants should accompany large-scale simulation experiments.

\noindent\textbf{Suggestion 6: (For researchers) Designing quantum-aware testing and validation frameworks for simulators.}
Our findings indicate that many simulator bugs arise from subtle semantic deviations, scaling assumptions, and cross-layer interactions that are rarely exercised by existing test suites. In particular, wrong outputs and silent failures frequently escape automated detection, while memory, overflow, and truncation errors manifest only under large or stress-inducing workloads. Researchers should develop quantum-aware testing methodologies that combine (1) property-based testing grounded in quantum invariants such as unitarity preservation, probability normalization, and circuit equivalence; (2) differential testing across independent simulator backends to expose semantic inconsistencies introduced by optimizations; (3) scaling-aware workload generation that systematically varies qubit counts, threading configurations, and backend targets to reveal overflow and resource-management defects; and (4) integration of memory-safety and concurrency analysis tools for C++/CUDA and language bindings. By unifying semantic validation, stress testing, and systems-level analysis, future research can reduce the reliance on user-reported failures and significantly strengthen simulator reliability before deployment.
\section{Threats to Validity}
\label{sec:threats}

\stitle{Internal Validity.}
Manual annotation introduces subjectivity in classifying root causes and manifestations, particularly for bugs spanning multiple categories. Two researchers independently labeled each issue and resolved disagreements through discussion using a shared annotation schema refined via pilot analysis. While dual annotation reduces bias, some residual subjectivity remains.

Distinguishing confirmed bugs from feature requests, documentation changes, or refactorings also poses a risk. We manually reviewed 484 candidate issues using explicit exclusion criteria and removed cases where fixes combined refactoring and bug resolution without clear attribution. This conservative filtering reduces noise but may exclude some valid defects.

\stitle{External Validity.}
Our dataset includes 12 widely used open-source quantum simulators from public GitHub repositories, spanning statevector, stabilizer, tensor-network, and GPU backends. Results may not generalize to proprietary simulators, closed hardware stacks, or projects with private issue trackers. We also include only issues resolved through accepted pull requests; latent, disputed, or silently fixed bugs are not captured. Thus, our findings reflect reported and fixed defects rather than the full universe of failures.

\stitle{Construct Validity.}
Our taxonomy of root causes and manifestations was developed iteratively and informed by prior empirical studies. However, categorization simplifies complex defects, and boundaries between categories (e.g., algorithmic vs.\ type-system errors) may be ambiguous. We assigned each issue to the dimension best representing its primary failure mechanism and allowed multiple tags when necessary.
Classifying algorithmic errors as quantum-specific versus classical is central to RQ3. We defined quantum-specific errors as those requiring knowledge of quantum mechanics, circuit semantics, or quantum-specific data structures to diagnose.

\stitle{Reliability.}
Root cause attribution depends on issue reports, pull request discussions, and code diffs. In older issues, reproduction details were sometimes missing. In such cases, we relied primarily on associated code changes. Inter-annotator agreement and public release of our dataset and annotation guidelines support reproducibility, though independent replication may yield minor differences in edge cases.

\section{Related Work}
\label{sec:related}

\subsection{Empirical Bug Studies in Quantum Software}

Empirical research on quantum software engineering has recently gained traction as the community seeks to understand the development, maintenance, and quality challenges of quantum computing tools.
Upadhyay et al.~\cite{Upadhyay25} conducted a large-scale empirical analysis of 157K issues across multiple quantum software repositories, revealing that 34\% of reported issues are specific to quantum computing concepts, highlighting the need for quantum-aware software engineering tools.
El Aoun et al.~\cite{El_aoun21} classified questions related to \gls*{qse} on Stack Exchange forums and found that a significant portion of the questions arise in the context of quantum execution, a stage in which simulators play a vital role.
Paltenghi and Pradel~\cite{Paltenghi22} analyzed 223 bugs from 18 open-source projects and provided insights into quantum-specific bugs and their recurring patterns.
Zhao et al.~\cite{zhao2023identifying} analyzed 36 bugs from Qiskit, offering insights on issues related to quantum programming.
Zhao et al.~\cite{zhao2023qml} examined 391 bugs across 22 repositories of nine popular quantum machine learning frameworks, finding that 28\% of bugs are quantum-specific and distilling a taxonomy of five symptoms and nine root causes.
Murillo et al.~\cite{Murillo25} discussed the overall roadmap and challenges of quantum software engineering, including the role of simulators in the software lifecycle.
These studies either treat simulators as part of a broader software stack or analyze a small number of bugs from a single framework.

\subsection{Empirical Bug Studies in Classical Software Infrastructure}

Empirical bug studies in non-quantum software systems offer useful methodological parallels and show that infrastructure-level failure patterns recur across domains.
Liu et al.~\cite{Liu2025} conducted a comprehensive study of bugs in the \code{rustc} compiler, manually reviewing 301 valid issues, and found that compiler bugs primarily arise from the type system and data-handling layers rather than the core compilation engine, and that existing testing tools struggle to detect non-crash errors.
Drosos et al.~\cite{Drosos2024When} analyzed 360 bugs in Infrastructure as Code systems including Ansible, Puppet, and Chef, characterizing how bugs manifest, their root causes, their reproduction requirements, and how they are fixed, finding that configuration and dependency issues dominate reliability problems.
Wang et al.~\cite{Wang2025} studied 92,542 crashes in publicly available Python machine learning notebooks, finding that over 40\% stem from API misuse and notebook-specific issues such as out-of-order execution, and that crashes concentrate in data preparation and model training stages.
Across these studies, a consistent pattern emerges: failures tend to concentrate in orchestration and integration layers rather than in the core computational engine, and critical bugs frequently escape automated testing.
This mirrors our findings on quantum simulators and reinforces the broader relevance of infrastructure-aware reliability research.

\noindent Our study extends prior work by focusing specifically on quantum simulators, filtering for issues with associated pull requests, and conducting manual qualitative analysis across 394 confirmed issues from 12 simulators to provide a structured view of the failure patterns surrounding quantum simulation infrastructure.
\section{Conclusions}
\label{sec:conclusions}
We presented an empirical study of bugs in quantum simulators based on 394 real-world issues from widely used open-source systems. We analyze how these failures arise, how they manifest, and how they are discovered in practice.
Our results show that many simulator failures originate from classical software infrastructure, including configuration, dependency compatibility, and memory management issues. At the same time, quantum-specific defects are concentrated in core execution components such as state evolution, gate operations, and noise modeling. We also find that incorrect results, including silent failures that produce plausible but wrong outputs, are common and difficult to detect. In addition, many severe failures are not caught by automated testing and are instead reported by users.
Based on these findings, we provide a set of practical suggestions for improving simulator reliability, including stronger validation of correctness, more systematic testing approaches, and better handling of classical infrastructure. We expect that this study can inform the community and motivate follow-up work on improving the robustness, testing, and evaluation of quantum simulation systems.

\balance
\bibliographystyle{plainnat}
\bibliography{sample-base}

\begin{thebibliography}{77}
\providecommand{\natexlab}[1]{#1}
\providecommand{\url}[1]{\texttt{#1}}
\expandafter\ifx\csname urlstyle\endcsname\relax
  \providecommand{\doi}[1]{doi: #1}\else
  \providecommand{\doi}{doi: \begingroup \urlstyle{rm}\Url}\fi

\bibitem[Ali et~al.(2021)Ali, Arcaini, Wang, and Yue]{Ali2021}
Shaukat Ali, Paolo Arcaini, Xinyi Wang, and Tao Yue.
\newblock Assessing the effectiveness of input and output coverage criteria for
  testing quantum programs.
\newblock In \emph{2021 14th IEEE Conference on Software Testing, Verification
  and Validation (ICST)}, pages 13--23, April 2021.
\newblock \doi{10.1109/ICST49551.2021.00014}.

\bibitem[{Amazon Web Services}(2024)]{aws_braket_testing}
{Amazon Web Services}.
\newblock Testing quantum programs with amazon braket.
\newblock
  \url{https://docs.aws.amazon.com/braket/latest/developerguide/braket-test.html},
  2024.
\newblock Accessed: 2026-02-04.

\bibitem[{Awesome Quantum Software Contributors}()]{qosf-repo}
{Awesome Quantum Software Contributors}.
\newblock qosf/awesome-quantum-software: {C}urated list of open-source quantum
  software projects.
\newblock \url{https://github.com/qosf/awesome-quantum-software/}.
\newblock Accessed: 2025-09-17.

\bibitem[Bergholm et~al.(2022)Bergholm, Izaac, Schuld, Gogolin, Ahmed, Ajith,
  Alam, Alonso-Linaje, AkashNarayanan, Asadi, Arrazola, Azad, Banning, Blank,
  Bromley, Cordier, Ceroni, Delgado, Matteo, Dusko, Garg, Guala, Hayes, Hill,
  Ijaz, Isacsson, Ittah, Jahangiri, Jain, Jiang, Khandelwal, Kottmann, Lang,
  Lee, Loke, Lowe, McKiernan, Meyer, Montañez-Barrera, Moyard, Niu, O'Riordan,
  Oud, Panigrahi, Park, Polatajko, Quesada, Roberts, Sá, Schoch, Shi, Shu,
  Sim, Singh, Strandberg, Soni, Száva, Thabet, Vargas-Hernández, Vincent,
  Vitucci, Weber, Wierichs, Wiersema, Willmann, Wong, Zhang, and
  Killoran]{pennylane}
Ville Bergholm, Josh Izaac, Maria Schuld, Christian Gogolin, Shahnawaz Ahmed,
  Vishnu Ajith, M.~Sohaib Alam, Guillermo Alonso-Linaje, B.~AkashNarayanan, Ali
  Asadi, Juan~Miguel Arrazola, Utkarsh Azad, Sam Banning, Carsten Blank,
  Thomas~R Bromley, Benjamin~A. Cordier, Jack Ceroni, Alain Delgado, Olivia~Di
  Matteo, Amintor Dusko, Tanya Garg, Diego Guala, Anthony Hayes, Ryan Hill,
  Aroosa Ijaz, Theodor Isacsson, David Ittah, Soran Jahangiri, Prateek Jain,
  Edward Jiang, Ankit Khandelwal, Korbinian Kottmann, Robert~A. Lang, Christina
  Lee, Thomas Loke, Angus Lowe, Keri McKiernan, Johannes~Jakob Meyer, J.~A.
  Montañez-Barrera, Romain Moyard, Zeyue Niu, Lee~James O'Riordan, Steven Oud,
  Ashish Panigrahi, Chae-Yeun Park, Daniel Polatajko, Nicolás Quesada, Chase
  Roberts, Nahum Sá, Isidor Schoch, Borun Shi, Shuli Shu, Sukin Sim, Arshpreet
  Singh, Ingrid Strandberg, Jay Soni, Antal Száva, Slimane Thabet, Rodrigo~A.
  Vargas-Hernández, Trevor Vincent, Nicola Vitucci, Maurice Weber, David
  Wierichs, Roeland Wiersema, Moritz Willmann, Vincent Wong, Shaoming Zhang,
  and Nathan Killoran.
\newblock Pennylane: Automatic differentiation of hybrid quantum-classical
  computations, 2022.
\newblock URL \url{https://arxiv.org/abs/1811.04968}.

\bibitem[Bidzhiev()]{PasqalEmulatorsIssue54}
Kemal Bidzhiev.
\newblock {B}istring sampling performance · {I}ssue \#54 ·
  pasqal-io/emulators --- github.com.
\newblock \url{https://github.com/pasqal-io/emulators/issues/54}.
\newblock Accessed: 2025-09-17.

\bibitem[Blaauw()]{QrackIssue234}
Aryan Blaauw.
\newblock 31 qubits or more crash benchmark.cpp with layers qengine and qfusion
  · {I}ssue \#234 · unitaryfoundation/qrack --- github.com.
\newblock \url{https://github.com/unitaryfoundation/qrack/issues/234}.
\newblock Accessed: 2025-09-17.

\bibitem[Bravyi et~al.(2022)Bravyi, Dial, Gambetta, Gil, and Nazario]{Bravyi22}
Sergey Bravyi, Oliver Dial, Jay~M. Gambetta, Darío Gil, and Zaira Nazario.
\newblock The future of quantum computing with superconducting qubits.
\newblock \emph{Journal of Applied Physics}, 132\penalty0 (16):\penalty0
  160902, 10 2022.
\newblock ISSN 0021-8979.
\newblock \doi{10.1063/5.0082975}.
\newblock URL \url{https://doi.org/10.1063/5.0082975}.

\bibitem[Chaliasos et~al.(2021)Chaliasos, Sotiropoulos, Drosos, Mitropoulos,
  Mitropoulos, and Spinellis]{chaliasos2021well}
Stefanos Chaliasos, Thodoris Sotiropoulos, Georgios-Petros Drosos, Charalambos
  Mitropoulos, Dimitris Mitropoulos, and Diomidis Spinellis.
\newblock Well-typed programs can go wrong: A study of typing-related bugs in
  jvm compilers.
\newblock \emph{Proceedings of the ACM on Programming Languages}, 5\penalty0
  (OOPSLA):\penalty0 1--30, 2021.

\bibitem[Chen()]{QiskitAerIssue755}
Hongxiang Chen.
\newblock {S}egmentation {F}ault during repeated execution of circuit using
  `density\_matric\_gpu` simulator) · {I}ssue \#755 · {Q}iskit/qiskit-aer ---
  github.com.
\newblock \url{https://github.com/Qiskit/qiskit-aer/issues/755}.
\newblock Accessed: 2025-09-17.

\bibitem[{cuQuantum Contributors}()]{cuQuantum}
{cuQuantum Contributors}.
\newblock {NVIDIA}/cu{Q}uantum: {H}ome for cu{Q}uantum {P}ython \& {NVIDIA}
  cu{Q}uantum {SDK} {C}++ samples.
\newblock \url{https://github.com/NVIDIA/cuQuantum/}.
\newblock Accessed: 2025-09-17.

\bibitem[Dahale()]{QulacsIssue632}
Gopal~Ramesh Dahale.
\newblock {E}mpty gradients when using `{Q}uantum{C}ircuit{O}ptimizer` ·
  {I}ssue \#632 · qulacs/qulacs --- github.com.
\newblock \url{https://github.com/qulacs/qulacs/issues/632}.
\newblock Accessed: 2025-09-17.

\bibitem[Developers(2025)]{cirq}
Cirq Developers.
\newblock \emph{Cirq}.
\newblock Zenodo, August 2025.
\newblock \doi{10.5281/ZENODO.4062499}.
\newblock URL \url{https://zenodo.org/doi/10.5281/zenodo.4062499}.

\bibitem[Doi()]{QiskitAerIssue1479}
Jun Doi.
\newblock {M}ulti-chunk parallelization is not applied to density\_matrix with
  noises · {I}ssue \#1479 · {Q}iskit/qiskit-aer --- github.com.
\newblock \url{https://github.com/Qiskit/qiskit-aer/issues/1479}.
\newblock Accessed: 2025-09-17.

\bibitem[Drosos et~al.(2024)Drosos, Sotiropoulos, Alexopoulos, Mitropoulos, and
  Su]{Drosos2024When}
Georgios-Petros Drosos, Thodoris Sotiropoulos, Georgios Alexopoulos, Dimitris
  Mitropoulos, and Zhendong Su.
\newblock When your infrastructure is a buggy program: Understanding faults in
  infrastructure as code ecosystems.
\newblock \emph{Proc. ACM Program. Lang.}, 8\penalty0 (OOPSLA2), October 2024.
\newblock \doi{10.1145/3689799}.
\newblock URL \url{https://doi.org/10.1145/3689799}.

\bibitem[El~aoun et~al.(2021)El~aoun, Li, Khomh, and Openja]{El_aoun21}
Mohamed~Raed El~aoun, Heng Li, Foutse Khomh, and Moses Openja.
\newblock Understanding quantum software engineering challenges an empirical
  study on stack exchange forums and github issues.
\newblock In \emph{2021 IEEE International Conference on Software Maintenance
  and Evolution (ICSME)}, pages 343--354, 2021.
\newblock \doi{10.1109/ICSME52107.2021.00037}.

\bibitem[Garrison()]{QiskitAerIssue2178}
Jim Garrison.
\newblock {D}eprecation{W}arning about {P}rovider · {I}ssue \#2178 ·
  {Q}iskit/qiskit-aer --- github.com.
\newblock \url{https://github.com/Qiskit/qiskit-aer/issues/2178}.
\newblock Accessed: 2025-09-17.

\bibitem[Giudice()]{StimIssue926}
Giacomo Giudice.
\newblock {G}etting `{F}lip{S}imulator` measurement/detector/observable flips
  with negative indexing fails · {I}ssue \#926 · quantumlib/{S}tim ---
  github.com.
\newblock \url{https://github.com/quantumlib/Stim/issues/926}.
\newblock Accessed: 2025-09-17.

\bibitem[Guerreschi et~al.(2020)Guerreschi, Hogaboam, Baruffa, and
  Sawaya]{intel-qs}
Gian~Giacomo Guerreschi, Justin Hogaboam, Fabio Baruffa, and Nicolas P~D
  Sawaya.
\newblock Intel quantum simulator: a cloud-ready high-performance simulator of
  quantum circuits.
\newblock \emph{Quantum Science and Technology}, 5\penalty0 (3):\penalty0
  034007, may 2020.
\newblock \doi{10.1088/2058-9565/ab8505}.
\newblock URL \url{https://doi.org/10.1088/2058-9565/ab8505}.

\bibitem[Heim()]{PyQtorchIssue35}
Niklas Heim.
\newblock `{H}am{E}vo` modules are mutating the input state · {I}ssue \#35 ·
  pasqal-io/pyqtorch --- github.com.
\newblock \url{https://github.com/pasqal-io/pyqtorch/issues/35}.
\newblock Accessed: 2025-09-17.

\bibitem[HiQsimulator()]{hiqsimulator}
HiQsimulator.
\newblock {M}ind{Q}uantum-{H}i{Q}/{H}i{Q}simulator: {A} high performance
  distributed quantum simulator --- github.com.
\newblock \url{https://github.com/MindQuantum-HiQ/HiQsimulator}.
\newblock Accessed: 2025-09-17.

\bibitem[Honarvar et~al.(2020)Honarvar, Mousavi, and Nagarajan]{Honarvar2020}
Shahin Honarvar, Mohammad~Reza Mousavi, and Rajagopal Nagarajan.
\newblock Property-based testing of quantum programs in q\#.
\newblock In \emph{Proceedings of the IEEE/ACM 42nd International Conference on
  Software Engineering Workshops}, ICSEW'20, page 430–435, New York, NY, USA,
  2020. Association for Computing Machinery.
\newblock ISBN 9781450379632.
\newblock \doi{10.1145/3387940.3391459}.
\newblock URL \url{https://doi.org/10.1145/3387940.3391459}.

\bibitem[@Hosseinberg()]{QsimIssue576}
@Hosseinberg.
\newblock {Q}{S}im{S}imulator.simulate\_expectation\_values for identity
  operators is wrong when the coefficient is other than one · {I}ssue \#576 ·
  quantumlib/qsim --- github.com.
\newblock \url{https://github.com/quantumlib/qsim/issues/576}.
\newblock Accessed: 2025-09-17.

\bibitem[{IBM Quantum}(2024)]{ibm_quantum_simulators}
{IBM Quantum}.
\newblock Using local simulators.
\newblock \url{https://quantum.cloud.ibm.com/docs/en/guides/local-simulators},
  2024.
\newblock Accessed: 2026-02-04.

\bibitem[Itoko()]{QiskitAerIssue1447}
Toshinari Itoko.
\newblock {R}elaxation{N}oise{P}ass creates noisy circuits that cannot be run
  directly on simulator for circuits with multi-qubit gates · {I}ssue \#1447
  · {Q}iskit/qiskit-aer --- github.com.
\newblock \url{https://github.com/Qiskit/qiskit-aer/issues/1447}.
\newblock Accessed: 2025-09-17.

\bibitem[@JamesB-1qbit()]{QulacsIssue303}
@JamesB-1qbit.
\newblock {M}emory leak from create\_quantum\_operator\_from\_openfermion\_text
  · {I}ssue \#303 · qulacs/qulacs --- github.com.
\newblock \url{https://github.com/qulacs/qulacs/issues/303}.
\newblock Accessed: 2025-09-17.

\bibitem[Javadi-Abhari()]{QiskitAerIssue1849}
Ali Javadi-Abhari.
\newblock bug in {E}stimator for parameterized dynamic circuits · {I}ssue
  \#1849 · {Q}iskit/qiskit-aer --- github.com.
\newblock \url{https://github.com/Qiskit/qiskit-aer/issues/1849}.
\newblock Accessed: 2025-09-17.

\bibitem[Javadi-Abhari et~al.(2024)Javadi-Abhari, Treinish, Krsulich, Wood,
  Lishman, Gacon, Martiel, Nation, Bishop, Cross, Johnson, and
  Gambetta]{qiskit}
Ali Javadi-Abhari, Matthew Treinish, Kevin Krsulich, Christopher~J. Wood, Jake
  Lishman, Julien Gacon, Simon Martiel, Paul~D. Nation, Lev~S. Bishop,
  Andrew~W. Cross, Blake~R. Johnson, and Jay~M. Gambetta.
\newblock Quantum computing with {Q}iskit, 2024.

\bibitem[@kosukemtr()]{QulacsIssue74}
@kosukemtr.
\newblock [{B}ug?] segmentation fault in python
  ({Q}uantum{C}ircuit.to\_string()) · {I}ssue \#74 · qulacs/qulacs ---
  github.com.
\newblock \url{https://github.com/qulacs/qulacs/issues/74}.
\newblock Accessed: 2025-09-17.

\bibitem[LaRose()]{QFlexIssue284}
Ryan LaRose.
\newblock {C}onfusing error message when copying rank-0 tensors · {I}ssue
  \#284 · s-mandra/qflex --- github.com.
\newblock \url{https://github.com/s-mandra/qflex/issues/284}.
\newblock Accessed: 2025-09-17.

\bibitem[Lattner and Adve(2004)]{llvm}
C.~Lattner and V.~Adve.
\newblock Llvm: a compilation framework for lifelong program analysis \&
  transformation.
\newblock In \emph{International Symposium on Code Generation and Optimization,
  2004. CGO 2004.}, pages 75--86, 2004.
\newblock \doi{10.1109/CGO.2004.1281665}.

\bibitem[Li et~al.(2019)Li, Ding, and Xie]{Li19}
Gushu Li, Yufei Ding, and Yuan Xie.
\newblock Tackling the qubit mapping problem for nisq-era quantum devices.
\newblock In \emph{Proceedings of the Twenty-Fourth International Conference on
  Architectural Support for Programming Languages and Operating Systems},
  ASPLOS '19, page 1001–1014, New York, NY, USA, 2019. Association for
  Computing Machinery.
\newblock ISBN 9781450362405.
\newblock \doi{10.1145/3297858.3304023}.
\newblock URL \url{https://doi.org/10.1145/3297858.3304023}.

\bibitem[Liu et~al.(2025)Liu, Feng, Ni, Li, Yin, Shi, Xu, and Su]{Liu2025}
Zixi Liu, Yang Feng, Yunbo Ni, Shaohua Li, Xizhe Yin, Qingkai Shi, Baowen Xu,
  and Zhendong Su.
\newblock An empirical study of bugs in the rustc compiler.
\newblock \emph{Proc. ACM Program. Lang.}, 9\penalty0 (OOPSLA2), October 2025.
\newblock \doi{10.1145/3763800}.
\newblock URL \url{https://doi.org/10.1145/3763800}.

\bibitem[MacIver et~al.(2019)MacIver, Hatfield-Dodds, and
  Contributors]{MacIver2019Hypothesis}
David MacIver, Zac Hatfield-Dodds, and Many Contributors.
\newblock Hypothesis: A new approach to property-based testing.
\newblock \emph{Journal of Open Source Software}, 4\penalty0 (43):\penalty0
  1891, 11 2019.
\newblock ISSN 2475-9066.
\newblock \doi{10.21105/joss.01891}.
\newblock URL \url{http://dx.doi.org/10.21105/joss.01891}.

\bibitem[Martin()]{QsimIssue283}
Orion Martin.
\newblock qtrajectory overwrites initial states · {I}ssue \#283 ·
  quantumlib/qsim --- github.com.
\newblock \url{https://github.com/quantumlib/qsim/issues/283}.
\newblock Accessed: 2025-09-17.

\bibitem[McKeeman(1998)]{MacKeeman1998Differential}
William~M. McKeeman.
\newblock Differential testing for software.
\newblock \emph{Digit. Tech. J.}, 10\penalty0 (1):\penalty0 100--107, 1998.

\bibitem[@merav aharoni()]{QiskitAerIssue997}
@merav aharoni.
\newblock {I}ncorrect result in {C}{C}{X} using {M}{P}{S} simulation method ·
  {I}ssue \#997 · {Q}iskit/qiskit-aer --- github.com.
\newblock \url{https://github.com/Qiskit/qiskit-aer/issues/997}.
\newblock Accessed: 2025-09-17.

\bibitem[Mishmash()]{QiskitAerIssue1308}
Ryan Mishmash.
\newblock {N}ondeterministic behavior in {M}{P}{S} when using multiple threads
  for sampling · {I}ssue \#1308 · {Q}iskit/qiskit-aer --- github.com.
\newblock \url{https://github.com/Qiskit/qiskit-aer/issues/1308}.
\newblock Accessed: 2025-09-17.

\bibitem[Moussa()]{PyQtorchIssue330}
Charles Moussa.
\newblock [{B}ug] {G}{P}{S}{R} with repeated parameter cannot be differentiated
  twice with shots · {I}ssue \#330 · pasqal-io/pyqtorch --- github.com.
\newblock \url{https://github.com/pasqal-io/pyqtorch/issues/330}.
\newblock Accessed: 2025-09-17.

\bibitem[Murillo et~al.(2025)Murillo, Garcia-Alonso, Moguel, Barzen, Leymann,
  Ali, Yue, Arcaini, P\'{e}rez-Castillo, Garc\'{\i}a-Rodr\'{\i}guez~de
  Guzm\'{a}n, Piattini, Ruiz-Cort\'{e}s, Brogi, Zhao, Miranskyy, and
  Wimmer]{Murillo25}
Juan~Manuel Murillo, Jose Garcia-Alonso, Enrique Moguel, Johanna Barzen, Frank
  Leymann, Shaukat Ali, Tao Yue, Paolo Arcaini, Ricardo P\'{e}rez-Castillo,
  Ignacio Garc\'{\i}a-Rodr\'{\i}guez~de Guzm\'{a}n, Mario Piattini, Antonio
  Ruiz-Cort\'{e}s, Antonio Brogi, Jianjun Zhao, Andriy Miranskyy, and Manuel
  Wimmer.
\newblock Quantum software engineering: Roadmap and challenges ahead.
\newblock \emph{ACM Trans. Softw. Eng. Methodol.}, 34\penalty0 (5), May 2025.
\newblock ISSN 1049-331X.
\newblock \doi{10.1145/3712002}.
\newblock URL \url{https://doi.org/10.1145/3712002}.

\bibitem[Nation({\natexlab{a}})]{QiskitAerIssue2249}
Paul Nation.
\newblock {A}er{E}stimator{V}2 fails to execute small circuits transpiled
  against larger backends · {I}ssue \#2249 · {Q}iskit/qiskit-aer ---
  github.com.
\newblock \url{https://github.com/Qiskit/qiskit-aer/issues/2249},
  {\natexlab{a}}.
\newblock Accessed: 2025-09-17.

\bibitem[Nation({\natexlab{b}})]{QiskitAerIssue741}
Paul Nation.
\newblock {P}ulse system model incosistently using dict based access · {I}ssue
  \#741 · {Q}iskit/qiskit-aer --- github.com.
\newblock \url{https://github.com/Qiskit/qiskit-aer/issues/741},
  {\natexlab{b}}.
\newblock Accessed: 2025-09-17.

\bibitem[Neeley({\natexlab{a}})]{QsimIssue327}
Matthew Neeley.
\newblock {Q}{S}im{S}imulator does not respect "idle" qubits · {I}ssue \#327
  · quantumlib/qsim --- github.com.
\newblock \url{https://github.com/quantumlib/qsim/issues/327}, {\natexlab{a}}.
\newblock Accessed: 2025-09-17.

\bibitem[Neeley({\natexlab{b}})]{QsimIssue482}
Matthew Neeley.
\newblock {Q}{S}im{S}imulator does not respect invert\_mask on measurements ·
  {I}ssue \#482 · quantumlib/qsim --- github.com.
\newblock \url{https://github.com/quantumlib/qsim/issues/482}, {\natexlab{b}}.
\newblock Accessed: 2025-09-17.

\bibitem[@NoureldinYosri()]{QsimIssue625}
@NoureldinYosri.
\newblock `{B}azel / {S}anitizer tests (msan)` {C}{I} failes due to {C}++ test
  failure · {I}ssue \#625 · quantumlib/qsim --- github.com.
\newblock \url{https://github.com/quantumlib/qsim/issues/625}.
\newblock Accessed: 2025-09-17.

\bibitem[O'Riordan()]{PennyLaneLightningIssue243}
Lee~James O'Riordan.
\newblock {U}pdate `pennylane.measure` usage due to recent changes in
  {P}enny{L}ane · {I}ssue \#243 · {P}enny{L}ane{A}{I}/pennylane-lightning ---
  github.com.
\newblock \url{https://github.com/PennyLaneAI/pennylane-lightning/issues/243}.
\newblock Accessed: 2025-09-17.

\bibitem[@oscarhiggott()]{StimIssue373}
@oscarhiggott.
\newblock macos sinter error: "module 'os' has no attribute
  'sched\_setaffinity'" · {I}ssue \#373 · quantumlib/{S}tim --- github.com.
\newblock \url{https://github.com/quantumlib/Stim/issues/373}.
\newblock Accessed: 2025-09-17.

\bibitem[Paltenghi and Pradel(2022)]{Paltenghi22}
Matteo Paltenghi and Michael Pradel.
\newblock Bugs in quantum computing platforms: an empirical study.
\newblock \emph{Proc. ACM Program. Lang.}, 6\penalty0 (OOPSLA1), April 2022.
\newblock \doi{10.1145/3527330}.
\newblock URL \url{https://doi.org/10.1145/3527330}.

\bibitem[Paltenghi and Pradel(2024)]{paltenghi2024surveytestinganalysisquantum}
Matteo Paltenghi and Michael Pradel.
\newblock A survey on testing and analysis of quantum software, 2024.
\newblock URL \url{https://arxiv.org/abs/2410.00650}.

\bibitem[Park()]{PennyLaneLightningIssue209}
Chae-Yeun Park.
\newblock {P}ytest fails on {W}indows · {I}ssue \#209 ·
  {P}enny{L}ane{A}{I}/pennylane-lightning --- github.com.
\newblock \url{https://github.com/PennyLaneAI/pennylane-lightning/issues/209}.
\newblock Accessed: 2025-09-17.

\bibitem[{Pasqal Emulators Contributors}()]{pasqal-emulators}
{Pasqal Emulators Contributors}.
\newblock pasqal-io/emulators: monorepo for pasqal torch based pulser backends.
\newblock \url{https://github.com/pasqal-io/emulators/}.
\newblock Accessed: 2025-09-17.

\bibitem[{PennyLane Lightning Contributors}()]{pennylane-lightning}
{PennyLane Lightning Contributors}.
\newblock {P}enny{L}ane{AI}/pennylane-lightning: {T}he {L}ightning plugin
  ecosystem provides fast quantum state-vector and tensor network simulators
  written in {C}++ for use with {P}enny{L}ane.
\newblock \url{https://github.com/PennyLaneAI/pennylane-lightning/}.
\newblock Accessed: 2025-09-17.

\bibitem[{PyQTorch Contributors}()]{pyqtorch}
{PyQTorch Contributors}.
\newblock pasqal-io/pyqtorch: {P}y{T}orch-based state vector and density matrix
  simulator.
\newblock \url{https://github.com/pasqal-io/pyqtorch/}.
\newblock Accessed: 2025-09-17.

\bibitem[{QFlex Contributors}()]{qflex}
{QFlex Contributors}.
\newblock s-mandra/qflex: {F}lexible {Q}uantum {C}ircuit {S}imulator (q{F}lex)
  implements an efficient tensor network, {CPU}-based simulator of large
  quantum circuits.
\newblock \url{https://github.com/s-mandra/qflex/}.
\newblock Accessed: 2025-09-17.

\bibitem[{Qiskit Contributors}()]{qiskit-aer}
{Qiskit Contributors}.
\newblock {Q}iskit/qiskit-aer: {A}er is a high performance simulator for
  quantum circuits that includes noise models.
\newblock \url{https://github.com/Qiskit/qiskit-aer/}.
\newblock Accessed: 2025-09-17.

\bibitem[{Q.js Contributors}()]{qjs}
{Q.js Contributors}.
\newblock stewdio/q.js: {Q}uantum computing in your browser.
\newblock \url{https://github.com/stewdio/q.js/}.
\newblock Accessed: 2025-09-17.

\bibitem[{Qrack Contributors}()]{qrack}
{Qrack Contributors}.
\newblock unitaryfoundation/qrack: {C}omprehensive, {G}{P}{U} accelerated
  framework for developing universal virtual quantum processors.
\newblock \url{github.com/unitaryfoundation/qrack/}.
\newblock Accessed: 2025-09-17.

\bibitem[{QSim Contributors}()]{qsim}
{QSim Contributors}.
\newblock quantumlib/qsim: {F}ast {C}++ and {P}ython library for state-vector
  simulation of quantum circuits.
\newblock \url{https://github.com/quantumlib/qsim/}.
\newblock Accessed: 2025-09-17.

\bibitem[{QuEST Contributors}()]{quest}
{QuEST Contributors}.
\newblock {Q}u{EST}-{K}it/{Q}u{EST}: {A} multithreaded, distributed,
  {GPU}-accelerated simulator of quantum computers.
\newblock \url{https://github.com/QuEST-Kit/QuEST/}.
\newblock Accessed: 2025-09-17.

\bibitem[{Quimb Contributors}()]{quimb}
{Quimb Contributors}.
\newblock jcmgray/quimb: {A} python library for quantum information and
  many-body calculations including tensor networks.
\newblock \url{https://github.com/jcmgray/quimb/}.
\newblock Accessed: 2025-09-17.

\bibitem[{Qulacs Contributors}()]{qulacs}
{Qulacs Contributors}.
\newblock qulacs/qulacs: {V}ariational {Q}uantum {C}ircuit {S}imulator for
  {Q}uantum {C}omputation {R}esearch.
\newblock \url{https://github.com/qulacs/qulacs/}.
\newblock Accessed: 2025-09-17.

\bibitem[{Quplexity Contributors}()]{quplexity}
{Quplexity Contributors}.
\newblock {M}r{G}illi/{Q}uplexity: {O}fficial repo of the very fast and
  lightweight modular library (or accelerator) for {Q}uantum {C}omputer
  {C}ircuit simulation.
\newblock \url{https://github.com/MrGilli/Quplexity/}.
\newblock Accessed: 2025-09-17.

\bibitem[{QVM Contributors}()]{qvm}
{QVM Contributors}.
\newblock quil-lang/qvm: {T}he high-performance and featureful {Q}uil
  simulator.
\newblock \url{https://github.com/quil-lang/qvm/}.
\newblock Accessed: 2025-09-17.

\bibitem[Retault()]{QiskitAerIssue1878}
Vi~Retault.
\newblock {S}egfault when simulating circuits with specific number of qubits ·
  {I}ssue \#1878 · {Q}iskit/qiskit-aer --- github.com.
\newblock \url{https://github.com/Qiskit/qiskit-aer/issues/1878}.
\newblock Accessed: 2025-09-17.

\bibitem[Schoenfeld()]{QiskitAerIssue416}
Zachary Schoenfeld.
\newblock {S}egmentation {F}ault on {P}ulse {S}imulator · {I}ssue \#416 ·
  {Q}iskit/qiskit-aer --- github.com.
\newblock \url{https://github.com/Qiskit/qiskit-aer/issues/416}.
\newblock Accessed: 2025-09-17.

\bibitem[Seaman(1999)]{kappa}
C.B. Seaman.
\newblock Qualitative methods in empirical studies of software engineering.
\newblock \emph{IEEE Transactions on Software Engineering}, 25\penalty0
  (4):\penalty0 557--572, 1999.
\newblock \doi{10.1109/32.799955}.

\bibitem[{Selene Contributors}()]{selene}
{Selene Contributors}.
\newblock {CQCL}/selene: {Q}uantinuum's emulator for hybrid quantum
  computation.
\newblock \url{https://github.com/CQCL/selene/}.
\newblock Accessed: 2025-09-17.

\bibitem[Serebryany and Iskhodzhanov(2009)]{Serebryany2009ThreadSanitizer}
Konstantin Serebryany and Timur Iskhodzhanov.
\newblock Threadsanitizer: data race detection in practice.
\newblock In \emph{Proceedings of the Workshop on Binary Instrumentation and
  Applications}, WBIA '09, page 62–71, New York, NY, USA, 2009. Association
  for Computing Machinery.
\newblock ISBN 9781605587936.
\newblock \doi{10.1145/1791194.1791203}.
\newblock URL \url{https://doi.org/10.1145/1791194.1791203}.

\bibitem[Serebryany et~al.(2012)Serebryany, Bruening, Potapenko, and
  Vyukov]{Konstantin2012AddressSanitizer}
Konstantin Serebryany, Derek Bruening, Alexander Potapenko, and Dmitry Vyukov.
\newblock Addresssanitizer: A fast address sanity checker.
\newblock In \emph{USENIX ATC 2012}, 2012.
\newblock URL
  \url{https://www.usenix.org/conference/usenixfederatedconferencesweek/addresssanitizer-fast-address-sanity-checker}.

\bibitem[{Stim Contributors}()]{stim}
{Stim Contributors}.
\newblock quantumlib/{S}tim: {A} fast stabilizer circuit library.
\newblock \url{https://github.com/quantumlib/Stim/}.
\newblock Accessed: 2025-09-17.

\bibitem[Strano()]{QrackIssue681}
Daniel Strano.
\newblock {Q}\# {S}{W}{A}{P}{T}wo{Q}ubits unit test failure · {I}ssue \#681 ·
  unitaryfoundation/qrack --- github.com.
\newblock \url{https://github.com/unitaryfoundation/qrack/issues/681}.
\newblock Accessed: 2025-09-17.

\bibitem[Upadhyay et~al.(2025)Upadhyay, Chhetri, Siddique, and
  Farooq]{Upadhyay25}
Krishna Upadhyay, Vinaik Chhetri, A.B. Siddique, and Umar Farooq.
\newblock { Analyzing the Evolution and Maintenance of Quantum Software
  Repositories }.
\newblock In \emph{2025 IEEE International Conference on Quantum Software
  (QSW)}, pages 173--184, Los Alamitos, CA, USA, July 2025. IEEE Computer
  Society.
\newblock \doi{10.1109/QSW67625.2025.00029}.
\newblock URL
  \url{https://doi.ieeecomputersociety.org/10.1109/QSW67625.2025.00029}.

\bibitem[Wang et~al.(2025)Wang, Meijer, Lopez, Nilsson, and Varro]{Wang2025}
Yiran Wang, Willem Meijer, Jose Antonio~Hernandez Lopez, Ulf Nilsson, and
  Daniel Varro.
\newblock { Why Do Machine Learning Notebooks Crash? An Empirical Study on
  Public Python Jupyter Notebooks }.
\newblock \emph{IEEE Transactions on Software Engineering}, 51\penalty0
  (07):\penalty0 2181--2196, July 2025.
\newblock ISSN 1939-3520.
\newblock \doi{10.1109/TSE.2025.3574500}.
\newblock URL
  \url{https://doi.ieeecomputersociety.org/10.1109/TSE.2025.3574500}.

\bibitem[Wood()]{QiskitAerIssue1193}
Christopher~J. Wood.
\newblock {I}ncorrect results when using cache\_blocking statevector simulator
  · {I}ssue \#1193 · {Q}iskit/qiskit-aer --- github.com.
\newblock \url{https://github.com/Qiskit/qiskit-aer/issues/1193}.
\newblock Accessed: 2025-09-17.

\bibitem[Xiong et~al.(2023)Xiong, Xu, Su, Sun, Wang, Wen, Pu, He, and
  Su]{xiong2023an}
Yiheng Xiong, Mengqian Xu, Ting Su, Jingling Sun, Jue Wang, He~Wen, Geguang Pu,
  Jifeng He, and Zhendong Su.
\newblock An empirical study of functional bugs in android apps.
\newblock In \emph{Proceedings of the 32nd ACM SIGSOFT International Symposium
  on Software Testing and Analysis}, ISSTA 2023, page 1319–1331, New York,
  NY, USA, 2023. Association for Computing Machinery.
\newblock ISBN 9798400702211.
\newblock \doi{10.1145/3597926.3598138}.
\newblock URL \url{https://doi.org/10.1145/3597926.3598138}.

\bibitem[Zhang()]{QulacsIssue314}
Zijian Zhang.
\newblock {E}xpectation value can be wrong when using {D}ensity{M}atrix
  ({S}evere bug) · {I}ssue \#314 · qulacs/qulacs --- github.com.
\newblock \url{https://github.com/qulacs/qulacs/issues/314}.
\newblock Accessed: 2025-09-17.

\bibitem[Zhao et~al.(2021)Zhao, Zhao, and Ma]{zhao2023identifying}
Pengzhan Zhao, Jianjun Zhao, and Lei Ma.
\newblock Identifying bug patterns in quantum programs.
\newblock In \emph{2021 IEEE/ACM 2nd International Workshop on Quantum Software
  Engineering (Q-SE)}, pages 16--21, 2021.
\newblock \doi{10.1109/Q-SE52541.2021.00011}.

\bibitem[Zhao et~al.(2023)Zhao, Wu, Luo, Li, and Zhao]{zhao2023qml}
Pengzhan Zhao, Xiongfei Wu, Junjie Luo, Zhuo Li, and Jianjun Zhao.
\newblock { An Empirical Study of Bugs in Quantum Machine Learning Frameworks
  }.
\newblock In \emph{2023 IEEE International Conference on Quantum Software
  (QSW)}, pages 68--75, Los Alamitos, CA, USA, July 2023. IEEE Computer
  Society.
\newblock \doi{10.1109/QSW59989.2023.00018}.
\newblock URL
  \url{https://doi.ieeecomputersociety.org/10.1109/QSW59989.2023.00018}.

\end{thebibliography}

\end{document}